\documentclass[12pt,preprint]{aastex}

\newcommand{\nn}{\nonumber}

\shorttitle{Particle acceleration around shocks III}
\shortauthors{Morlino, Vietri & Blasi}

\begin{document}

\title{Particle acceleration at shock waves moving at arbitrary speed:\\
   the case of large scale magnetic field and anisotropic scattering}

\author{G. Morlino}
\affil{Dipartimento di Fisica, Universit\`a di Pisa,
    Pisa, Italy}

\author{P. Blasi}
\affil{INAF/Osservatorio Astronomico di Arcetri, Firenze, Italy}

\and

\author{M. Vietri}
\affil{Scuola Normale Superiore, Pisa, Italy}

\begin{abstract}

A mathematical approach to investigate particle acceleration at shock
waves moving at arbitrary speed in a medium with arbitrary scattering
properties was first discussed in \cite{vie03,bla05}. We use this
method and somewhat extend it in order to include the effect of a large
scale magnetic field in the upstream plasma, with arbitrary
orientation with respect to the direction of motion of the shock. We
also use this approach to investigate the effects of anisotropic
scattering on spectra and anisotropies of the distribution function
of the accelerated particles.
\end{abstract}

\keywords{cosmic rays -- shock waves}

\section{Introduction}

The theory of particle acceleration at shock fronts moving with
arbitrary speeds (from newtonian to ultra-relativistic) can be
formulated in a simple and exact form \cite{vie03,fvy}, at least in
the so-called {\it test particle} limit, which neglects the
dynamical reaction of accelerated particles on the shock. In this
framework, all the basic physical ingredients can be taken into
account in an exact way, with special reference to the type of
scattering that is responsible for the particles to return to the
shock front from the upstream and downstream plasmas. The
information about scattering is introduced in the problem through a
function $w(\mu,\mu')$ which expresses the probability per unit
length that a particle moving in the direction $\mu'$ is scattered
to a new direction $\mu$. It is worth stressing that $w$ can have a
different functional form in the upstream and downstream plasmas, in
particular in the case of relativistic shocks.

The repeated scatterings of the particles lead eventually to return to the shock
front, as described in terms of the conditional probability
$P_u(\mu_0,\mu)$ ($P_d(\mu_0,\mu)$) that a particle entering the
upstream (downstream) plasma in the direction $\mu_0$, returns to
the shock and crosses it in the direction of the downstream (upstream)
plasma in the direction $\mu$. The mathematical method adopted to
calculate the two very important functions $P_u$ and $P_d$
based upon the knowledge of the elementary scattering function $w$ was
described in detail in \cite{bla05}, and is based on solving two
non-linear integral-differential equations in the two independent
coordinates $\mu_0$ and $\mu$.

\cite{vie03} showed on very general grounds that the spectrum of accelerated
particles is a power law for all momenta exceeding the injection momentum.
The slope of such power law and the anisotropy pattern of the
accelerated particles near the shock front are fully determined by the
conditional probabilities $P_d$ and $P_u$ and by the equation of state
of the downstream plasma. Particle acceleration at shock fronts has been
previously investigated through different methods, both semi-analytical
(see for example \cite{kir87,gal99,kir00,ach01}) and numerical, by using
Monte Carlo simulations (\textit{e.g.} \cite{bed98,lem03,nie04,lem05}).
The theory of particle acceleration developed by \cite{vie03} and \cite{bla05}
has been checked versus several of these calculations existing in the
literature, both in the case of non relativistic shocks and for
relativistic shocks, and assuming small as well as large pitch angle
isotropic scattering (see \cite{bla05} for an extensive discussion of
these results).

In this paper we extend the application of this new theoretical
framework to two new interesting situations: 1) presence of a coherent
large scale magnetic field in the upstream fluid; 2) anisotropic scattering.
In both cases we calculate the spectrum of accelerated particles and
the distribution in pitch angle (upstream and downstream) for shock
fronts moving with arbitrary velocity. The results of point 1) are
compared with those obtained in \cite{ach01}, carried out for a
parallel ultra-relativistic shock.

The paper has been inspired by the need to address several points of
phenomenological relevance. As far as relativistic shocks are
concerned, it was understood that the return of the particles to the
shock surface from the upstream region can be warranted even in the
absence of scattering, provided the background magnetic field is at
an angle with the shock normal (e.g. \cite{ach01}). This is due to
the fact that the shock and the accelerated particles remain
spatially close and regular deflection takes place before particles
can experience the complex, possibly turbulent structure of the
upstream magnetic field. This implies that the calculation of the
spectrum of the accelerated particles cannot be calculated using a
formalism based on the assumption of pitch angle diffusion, as in
the vast majority of the existing literature.

In the downstream region, the motion of the shock is always
quasi-newtonian, even when the shock moves at ultra-relativistic
speeds. This implies that the propagation of the particles is
generally well described by (small or large) pitch angle
scattering. However, the turbulent structure of the magnetic field,
responsible for the scattering, is likely to have an anisotropic
structure and be therefore responsible for anisotropic scattering. In
fact, even in the case of isotropic turbulence, the scattering can
determine an anisotropic pattern of particle scattering. It follows
that a determination of the spectrum able to take into account these
potentially important situations is very important.

The outline of the paper is the following: in \S \ref{sec:exact}
we briefly summarize the theoretical framework introduced in
\cite{vie03} and \cite{bla05}. In \S \ref{sec:regular} we consider
in detail the case of a large scale magnetic field in
the upstream frame and no scattering of the particles. The scattering
is assumed to be isotropic in the downstream plasma. In \S
\ref{sec:ani} we introduce the possibility of anisotropic
scattering in both upstream and downstream plasmas. We summarize in
\S \ref{sec:conclusions}.

\section{An exact solution for the accelerated particles in arbitrary
  conditions: a summary} \label{sec:exact}

In this section we summarize the main characteristics of the theory of
particle acceleration developed by \cite{vie03} and \cite{bla05}. The
reader is referred to this previous work for further
details. The power of this novel approach is in its generality: it
provides an exact solution for the spectrum of the accelerated
particles and at the same time the distribution in pitch angle that
the particles acquire due to scattering in the upstream and downstream
fluids. This mathematical approach is applicable without restrictions
on the velocity of the fluid speeds (from newtonian to
ultra-relativistic) and irrespective of the scattering properties of
the background plasmas (small as well as large angle scattering,
isotropic or anisotropic scattering). The only condition which is
necessary for the theory to work is common to most if not all other
semi-analytical approaches existing in the literature, namely that the
acceleration must take place in the test particles regime: no
dynamical reaction is currently introduced in the calculations.
As a consequence, the shock is assumed to conserve its strength during
the acceleration time, and the acceleration is assumed to
have reached a stationary regime.

The directions of motion of the particles in the downstream and
upstream frames are identified through the cosine of their pitch
angles, all evaluated in the comoving frames of the fluids that they
refer to. The direction of motion of the shock, identified as the $z$
axis, is assumed to be oriented from upstream to downstream, following
the direction of motion of the fluid in the shock frame ($\mu=1$
corresponds to particles moving toward the downstream section).

The transport equation for the particle distribution function $g$,
as obtained in \cite{vie03} in a relativistically covariant
derivation reads

\begin{equation} \label{sez2eq_transport_eq}
\gamma (u+\mu) {\partial g \over \partial z} =
       \int \!\!\! \int \left[{-W(\mu',\mu,\phi',\phi)
       g(\mu,\phi) + W(\mu,\mu',\phi,\phi') g(\mu',\phi')} \right]
        d\mu' d\phi'
       + \omega {\partial g \over \partial \tilde\phi} ,
\end{equation}
in which both scattering and regular deflection in a large scale
magnetic field are taken into account.

Here all quantities are written in the fluid frame, with the exception
of the spatial coordinate $z$, the distance from the shock along the
shock normal, which is measured in the shock frame. $u$ and $\gamma$
are, respectively, the velocity and the Lorentz factor of the fluid
with respect to the shock. $\theta$ and $\phi$ are the polar
coordinates of particles in momentum space, measured with respect to
the shock normal, while $\tilde\phi$ is the longitudinal angle around
the magnetic field direction. As usual $\mu=\cos\theta$ and
$\omega = e B/E$ is the particle Larmor frequency. $W(\mu,\mu',\phi,\phi')$
is the scattering probability per unit length, namely the probability
that a particle moving in the direction ($\mu',\phi'$) is scattered
to a direction ($\mu,\phi$) after travelling a unit length.

An important simplification of eq. (\ref{sez2eq_transport_eq}) occurs
when an axial symmetry is assumed. In this case the scattering
probability depends only on $\Delta \equiv \phi-\phi'$ and the
large scale magnetic field can be either zero or different from zero but
parallel to the shock normal. In both cases it is straightforward to
integrate eq. (\ref{sez2eq_transport_eq}) over $\phi$: the two-dimensional
integral on the right-hand side simplifies to an integral in
one dimension, while the term $\omega (\partial g / \partial
\tilde{\phi})$ disappears.

These simplifications lead to
\begin{equation} \label{sez2eq_transport_eq2}
\gamma (u+\mu) {\partial g \over \partial z} =
       \int \left[{-w(\mu',\mu) g(\mu) + w(\mu,\mu') g(\mu')} \right] d\mu' ,
\end{equation}
where
\begin{eqnarray} \label{sez2eq_scat_prob}
w(\mu,\mu') &\equiv&  \int W(\mu,\mu',\Delta) \, d\Delta  \qquad {\rm and} \nn
\\
g(\mu)      &\equiv& {1 \over 2\pi} \int g(\mu,\phi) \, d\phi \, .  \nn
\end{eqnarray}

The physical ingredients are all contained in the two conditional
probabilities $P_u(\mu_\circ,\mu)$ and $P_d(\mu_\circ,\mu)$: these two
functions provide respectively the probability that a particle
entering the upstream (downstream) plasma along a direction $\mu_0$
exits it along a direction $\mu$. In the absence of large scale
coherent magnetic fields, the two functions $P_u(\mu_\circ,\mu)$ and
$P_d(\mu_\circ,\mu)$ were defined through a set of two
integral-differential non linear equations by \cite{bla05}. We
report these equations here for completeness:

\begin{eqnarray}\label{sez2eq_Pu}
P_u(\mu_\circ,\mu)\left(\frac{d(\mu_\circ)}{u+\mu_\circ}-\frac{d(\mu)}{u+\mu}
\right)= \frac{w(\mu,\mu_\circ)}{u+\mu_\circ} - \int_{-u}^1
d\!\mu' \frac{w(\mu,\mu')P_u(\mu_\circ,\mu')}{u+\mu'}+ \nonumber\\
\int_{-1}^{-u}d\!\mu'\frac{w(\mu',\mu_\circ)P_u(\mu',\mu)}{u+\mu_\circ}
-\int_{-1}^{-u} d\!\mu' P_u(\mu',\mu)\int_{-u}^1
d\!\mu''\frac{w(\mu',\mu'') P_u(\mu_\circ,\mu'')}{u+\mu''},
\label{eq:Pu}
\end{eqnarray}

\begin{eqnarray}\label{sez2eq_Pd}
P_d(\mu_\circ,\mu)
\left(\frac{d(\mu_\circ)}{u+\mu_\circ}-\frac{d(\mu)}{u+\mu}
\right) = \frac{w(\mu,\mu_\circ)}{u+\mu_\circ} + \int_{-u}^1
d\!\mu' \frac{P_d(\mu',\mu) w(\mu',\mu_\circ)}{u+\mu_\circ} -
\nonumber \\
 \int_{-1}^{-u} d\!\mu' \frac{P_d(\mu_\circ,\mu')
w(\mu,\mu')}{u+\mu'}  - \int_{-u}^1 d\!\mu' P_d(\mu',\mu)
\int_{-1}^{-u} d\!\mu'' \frac{w(\mu',\mu'')
P_d(\mu_\circ,\mu'')}{u+\mu''}.
\label{eq:Pd}
\end{eqnarray}

In the equations above we used:
\begin{equation}
d(\mu) \equiv \int_{-1}^{+1} w(\mu',\mu) d\!\mu'\;,
\end{equation}
which is unity by definition.

It is worth stressing that Eq. (\ref{sez2eq_Pu}) provides automatically
the correct normalization for the return probability from upstream:
$\int_{-u}^1 d\mu'P_u(\mu_\circ,\mu') = 1$, independent of the entrance
angle $\mu_\circ$.
In \S \ref{sec:regular} we will generalize the method to include the possibility of
deflection by large scale magnetic fields, which is one of the
achievements of this work. In that case we will show that the
return probability from upstream is no longer bound to be unity,
due to the escape of particles from the upstream region.


The procedure for the calculation of the slope of the spectrum of
accelerated particles, as found by \cite{vie03} and \cite{bla05},
is as follows: for
a given Lorentz factor of the shock ($\gamma_{s}$), the velocity
of the upstream fluid $u=\beta_{s}$ is calculated. The velocity $u_d$
of the downstream fluid is found from the usual jump conditions at the
shock and through the adoption of an equation of state for the
downstream fluid.

Once the two functions $P_u$ and $P_d$ have been calculated, the slope
of the spectrum, as discussed in \cite{vie03}, is given by the
solution of the integral equation:

\begin{equation}
(u_d+\mu) g(\mu) = \int_{-u_d}^1 d\xi Q^T (\xi,\mu) (u_d+\xi) g(\xi),
\label{sez2eq_ang-dis}
\end{equation}
where
\begin{equation}
Q^T(\xi,\mu) = \int_{-1}^{-u_d} d\nu P_u(\nu,\mu) P_d(\xi,\nu)
\left(\frac{1-u_{rel}\mu}{1-u_{rel}\nu}\right)^{3-s}.
\label{sez2eq_Q^T}
\end{equation}

Here $u_{rel}=\frac{u-u_d}{1-u u_d}$ is the relative velocity
between the upstream and downstream fluids and $g(\mu)$ is the
angular part of the distribution function of the accelerated
particles, which contains all the information about the
anisotropy. Note that in Eq. (\ref{sez2eq_Q^T}) all variables and
functions are evaluated in the downstream frame, while the $P_u$
calculated through Eq. (\ref{eq:Pu}) is in the frame comoving
with the (upstream) fluid.
The $P_u$ that need to be used in Eq. (\ref{sez2eq_Q^T}) is therefore
$$ P_u(\nu,\mu) = P_u(\tilde \nu , \tilde \mu) \frac{d \tilde
\mu}{d\mu}= P_u(\tilde \nu , \tilde \mu)
\left[\frac{1-u_{rel}^2}{(1-u_{rel}\mu)^2} \right]. $$ The
solution for the slope $s$ of the spectrum is found by solving Eq.
(\ref{sez2eq_ang-dis}). In general, this equation has no solution but for
one value of $s$. Finding this value provides not only the slope
of the spectrum but also the angular distribution function $g(\mu)$.

\subsection{The special case of isotropic scattering}\label{sec:special}

No assumption has been introduced so far about the scattering
processes that determine the motion of the particles in the upstream
and downstream plasmas, with the exception of the axial symmetry of the
function $W(\mu,\mu',\phi,\phi')$.

A special case of this symmetric situation is that of isotropic
scattering, that takes place when the scattering probability $W$ only
depends upon the deflection angle $\Theta$, related to the initial and
final directions through
\begin{equation}
\cos\Theta \equiv \mu\mu' + \sqrt{1-\mu^2} \sqrt{1-\mu'^2} \cos(\phi-\phi').
\end{equation}

Among the many functional forms that correspond physically to
isotropic scattering, the simplest one is

\begin{equation} \label{sez2eq_w_iso_1}
W(\mu,\mu',\phi,\phi') = W(\cos\Theta) =
{1\over \sigma} e^{-{1-\cos\Theta \over \sigma}} \, ,
\end{equation}
where $\sigma$ is the mean scattering angle. Integration of eq.
(\ref{sez2eq_w_iso_1}) over $\phi - \phi'$ leads to
\begin{equation} \label{sez2eq_w_iso}
w(\mu,\mu') =
{1\over \sigma} e^{-{1-\mu\mu' \over \sigma}}
I_0\left( {\sqrt{1-\mu^2} \sqrt{1-\mu'^2} \over \sigma}\right) \, ,
\end{equation}
with $I_0(x)$ the Bessel function of order
0. Eq. (\ref{sez2eq_w_iso_1}), first introduced in \cite{bla05},
naturally satisfies the requirement of being symmetric under rotations
around the normal to the shock surface. In the limit $\sigma \ll 1$
this function becomes a Dirac Delta function, strongly peaked around
the forward direction, corresponding to isotropic Small Pitch Angle
Scattering (SPAS). For the opposite limit, that is $\sigma \gg 1$,
$w$ becomes flat and corresponds to the case of isotropic Large Angle
Scattering (LAS). In \S \ref{sec:model_ani} we will modify this functional
form to introduce the possibility of anisotropic scattering.

\section{Deflection by a regular magnetic field in the upstream region}
\label{sec:regular}

It is well known that particle acceleration at a shock front with
parallel magnetic field without scattering centers does not work.
This magnetic scattering may be self-generated by the same particles,
but the process of generation depends on the conditions in specific
astrophysical environments. The case in which a regular magnetic field
not parallel to the shock normal is present in the upstream fluid is
quite interesting in that it allows for the return of the particles to
the shock front even in the absence of scattering. In this section we
investigate in detail the process of acceleration at shocks with
arbitrary velocity when only a regular large scale magnetic field is
present upstream (no scattering). We assume that enough turbulence is
instead present in the downstream plasma to guarantee magnetic
scattering of the particles.

There are two main differences introduced by this situation when
compared with the standard case considered in the previous section:

\begin{itemize}
\item \textit{(a)} Particle motion in the upstream region is
  deterministic: the stochasticity introduced by the interaction with
  scattering centers is assumed to be absent. This requires a new
  determination of the return probability $P_u$ introduced above.

\item \textit{(b)} The presence of regular magnetic field with
  arbitrary orientation breaks the axial symmetry around the shock
  normal. This, in principle, would force us to treat the problem
  in the four angular variables $\mu_\circ, \phi_\circ, \mu$ and $\phi$.
\end{itemize}

In the following we will show how addressing point \textit{(a)} in
fact solves point \textit{(b)} as well.

\subsection{Upstream return probability}\label{sec:upP}

Let us Consider a particle entering upstream in the direction
identified by the two angles $\mu_\circ$ and $\phi_\circ$, and
returning to the shock along the direction identified by $\mu$ and $\phi$.
Since the motion of the particle is deterministic, the return direction is
completely defined by the incoming coordinates, and we can write in
full generality:
\begin{equation}\label{sez3eq_Pu_B_def}
P_u(\mu_\circ,\phi_\circ;\mu,\phi) = {(2 \pi)}^{-1}
\delta \left(\mu-\mu_1(\mu_\circ,\phi_\circ) \right) \,
\delta\left( \phi-\phi_1(\mu_\circ,\phi_\circ) \right) \, ,
\end{equation}
where $\mu_1$ and $\phi_1$ are obtained from the solution of the
equation of motion, as discussed below. One can see that $P_u$ is
effectively a function of only two variables.

In order to apply the same mathematical procedure introduced in \S \ref{sec:exact},
we need to write $P_u$ as a function of azimuthal angles
only. Therefore we use the properties of the delta function in
$\delta (\mu-\mu_1(\mu_\circ,\phi_\circ))$, to write:
\begin{eqnarray}\label{sez3eq_Pu_B_2}
P_u(\mu_\circ,\phi_\circ;\mu,\phi) &=&  {1 \over 2\pi} \left|
{\partial \phi_\circ (\mu_\circ,\mu)} \over \partial \mu \right|\,
\delta \left(\phi_\circ -\bar{\phi_\circ} \right) \, \delta
\left( \phi-\phi_1 \right) \nn  \\
&\equiv& P_u(\mu_\circ,\mu) \, \delta \left(\phi_\circ -
\bar{\phi_\circ} \right) \, \delta\left( \phi-\phi_1 \right) \, .
\end{eqnarray}

We now show that $P_u(\mu_\circ,\mu)$, as defined by Eq.
(\ref{sez3eq_Pu_B_2}), is exactly the function to be used in
eq. (\ref{sez2eq_Q^T}). This is easily shown by writing the fluxes of
particles ingoing and outgoing the upstream plasma:

\begin{equation}
J_+(\mu,\phi) = \int_{-1}^{-u} d\mu' \int_0^{2\pi} d\phi' \,
P_u(\mu',\phi',\mu,\phi) J_-(\mu',\phi') \, , \label{sez3eq_flux_+}
\end{equation}
%
which, when integrated over $\phi$, yields
\begin{equation}
J_+(\mu) \equiv \int d\phi J_+(\mu,\phi)  = \int_{-1}^{-u} d\mu'  \,
P_u(\mu',\mu)  J_-(\mu') \, ,
\end{equation}
where we assumed that $J_-$ is independent of $\phi$.  This is
exactly the same relationship as was used in \cite{vie03}, and
proves our point that the system may, in the average, still be
treated as if it were symmetric about the shock normal.

The key assumption here is that the flux crossing back into the
upstream region from the downstream one, $J_-$, be independent of
the azimuthal angle $\phi$. This is of course true in the Newtonian
regime, because there the residence time for all particles diverges,
and there is time for deflections to effectively erase anisotropies
in the $\phi$ direction. But this must be true {\it a fortiori} in
the relativistic regime, when one considers that the properties of
scattering are of course still the same as in the Newtonian regime,
while the surface to be recrossed, {\it i.e.}, the shock, is running
away from the particles at a speed that becomes, asymptotically, a
fair fraction of the particles' speed. So, while not exactly true,
the independence of $J_-$ from $\phi$ is at least a good
approximation.

In order to write $P_u(\mu_\circ,\mu)$ in a more explicit way, we need
to solve the equation of motion of the particles, namely find the
direction at which the particles re-cross the shock front as a
function of the incoming direction. Particles move
following a helicoidal trajectory around the magnetic field
direction, indicated here as $\tilde z$. The problem is simplest
if expressed in the frame  $\tilde O$ comoving with the upstream
fluid but with the polar axis coincident with $\tilde z$. We mark
with a tilde all quantities expressed in this frame. The equations
of motion in the frame $\tilde O$ are:
\begin{eqnarray} \label{sez3eq_tilde_motion}
\tilde\mu(t)  &=& \tilde{\mu}_{\circ}  \, , \\
\tilde\phi(t) &=& \tilde{\phi}_{\circ} + \omega t \, ,
\end{eqnarray}
where $t$ is time and $\omega$ is the Larmor frequency. The particles
re-cross the shock when $z_{\rm particle}(t)  = z_{\rm
  shock}(t)$. This condition expressed in the  frame $\tilde O$ reads
\begin{equation} \label{sez3eq_shock_encounter}
\sin(\omega \, t +\tilde{\phi}_{\circ}) -\sin\tilde{\phi}_{\circ} =
{\tilde{\mu}_{\circ} \, \cos\alpha  +\beta_s \over \sin\alpha \:
\sin\tilde{\theta}_\circ} \: \omega t \, ,
\end{equation}
where $\alpha$ is the angle between the shock normal $z$ and the
magnetic field direction $\tilde z$. The solution of Eq.
(\ref{sez3eq_shock_encounter}) gives the upstream residence time
$t^*$ of the particles, to be evaluated numerically.

\begin{figure}
\includegraphics[angle=0,scale=.8]{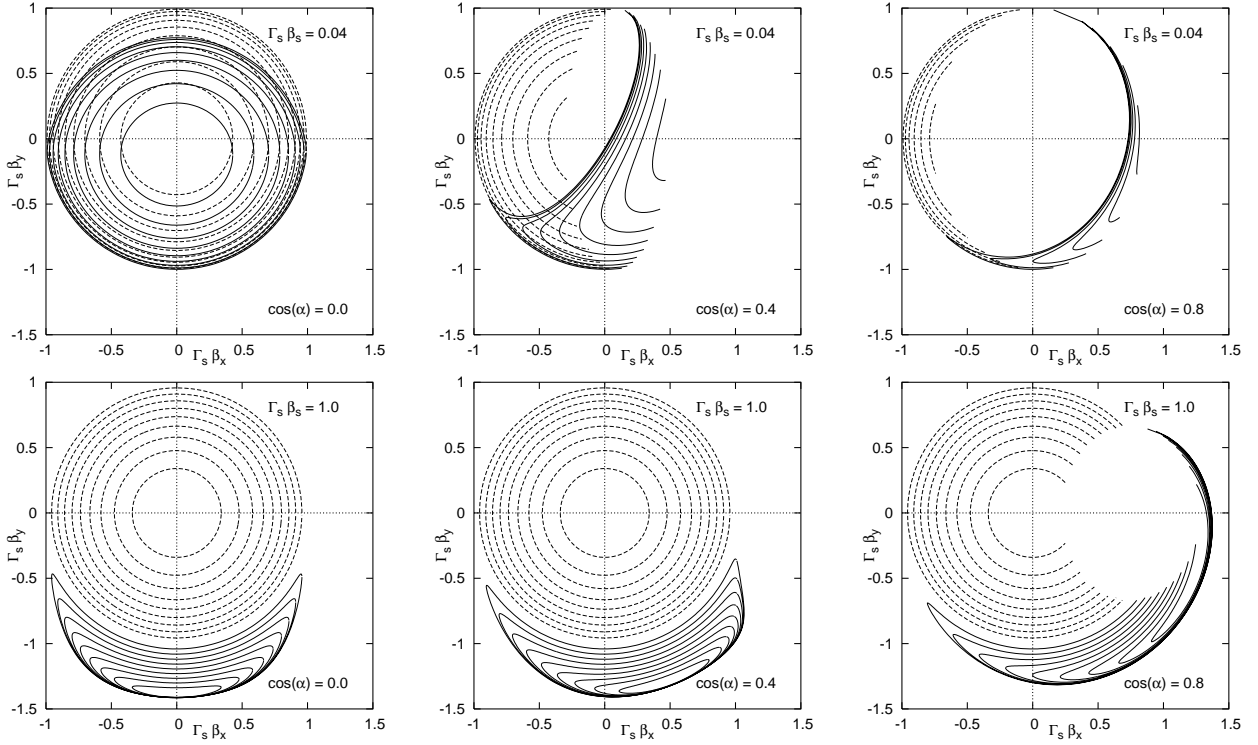}
\caption{Location of the particles entering the upstream region
(dashed lines) and returning to the downstream region (solid lines)
after being deflected by the magnetic field upstream. The directions
are plotted in the plane $\Gamma_s \beta_{p,x}$ - $\Gamma_s
\beta_{p,y}$. $\beta_{p,x}$ and $\beta_{p,y}$ are the components of
the particle velocity along the $x$ and the $y$ axis respectively.
The origin corresponds to particles entering along the shock normal.
Circles correspond to particles having constant $\mu_\circ$.
The upper panels refer to $\Gamma_s\beta_s=0.04$. The lower panel
refer to $\Gamma_s\beta_s=1.0$. In both cases we show the effects of
three different orientations of the magnetic field (from left to
right $\cos\alpha=0.0,0.4,0.8$).
The presence of an empty region when $\cos\alpha>\beta_s$ is due to
particle leakage from upstream [Compare with fig.1 in \cite{ach01}.]
\label{sez3fig_particle_in_out}}
\end{figure}

The angles that identify the re-crossing direction, as functions of
the residence time, are
\begin{eqnarray}
\tilde\mu_1  &=& \tilde{\mu}_{\circ} \, , \qquad {\rm and}  \\
\tilde\phi_1 &=& \tilde{\phi}_{\circ} +
\omega t^*(\tilde\mu_\circ,\tilde\phi_\circ) \, .
\end{eqnarray}
A rotation by the angle $\alpha$ provides us with the re-crossing
coordinates $\mu_1$ and $\phi_1$ in the fluid frame. At this point
the Jacobian in Eq. (\ref{sez3eq_Pu_B_2}) can be calculated, although
some care is needed because this Jacobian is not a single valued
function: for each pair ($\mu_\circ,\mu$) the Jacobian has two
values. This degeneracy arises because of the substitution of $\phi_\circ$
with $\mu$, since each $\mu$ corresponds in general to two possible
values of $\phi_\circ$. This is clear from fig. \ref{sez3fig_particle_in_out},
where we show some examples of solutions: the directions of entrance
and escape from the upstream fluid are plotted for different values
of the shock speed and for different orientations of the large scale
magnetic field.

Eq. (\ref{sez3eq_shock_encounter}) admits a solution $t^*>0$ only
if the two following conditions are fulfilled:
\begin{itemize}
\item[{\it i)}] the initial velocity of a particle along the shock
normal must be larger than the shock speed (otherwise the particle is
prevented from crossing the shock to start with). This implies:
\begin{equation} \label{sez3eq_limit1}
\mu_\circ < -\beta_s \, .
\end{equation}

\item[{\it ii)}] The particle velocity along the shock normal has to be less
than the shock speed, namely
\begin{equation} \label{sez3eq_limit2}
\tilde\mu_\circ \, \cos\alpha>-\beta_s \, .
\end{equation}
\end{itemize}

Particles not satisfying this last condition escape the shock region
towards upstream infinity, a situation which is not realized in
the case of scattering considered in \S \ref{sec:exact}. This escape process
occurs
only for $\cos\alpha>-\beta_s$, and results in the loss of particles
having the entrance pitch angles cosine exceeding
$\mu_{\rm min}(\mu_\circ,\phi_\circ)$. In fact for
$\cos\alpha<-\beta_s$, $\mu_{\rm min} = {\rm const}=-1$ and all
particles eventually re-cross the shock.

When the particles are allowed to escape upstream, the acceleration is
clearly expected to become less efficient and give rise to softer
spectra of the accelerated particles (see \S \ref{sec:sp_ani}).

Putting together all of the above, we can finally write the upstream
conditional probability as
\begin{equation}\label{sez3eq_Pu_B_int}
P_u(\mu_\circ,\mu) = {1 \over 2\pi} \sum_{i=1,2}
\left| {\partial \phi_\circ} \over \partial \mu \right|_i
\theta(-\mu_\circ-\beta_s) \, \theta(\mu_\circ-\mu_{\rm min}(\mu_\circ,\mu))\, ,
\end{equation}
where the sum is extended over the two branches of the Jacobian.

For $\cos\alpha<-\beta_s$ the particles always return to the
shock front and this forces the return probability to be unity
when integrated over all outgoing directions:
\begin{equation}
\int_{-u}^{1}d\mu \, P_u(\mu_\circ,\mu) = 1 \, .
\end{equation}
This integral condition is trivially satisfied by
Eq. (\ref{sez3eq_Pu_B_int}) and is used as a check for $P_u$ after its
numerical computation.

\begin{figure}
\includegraphics[angle=0,scale=.62]{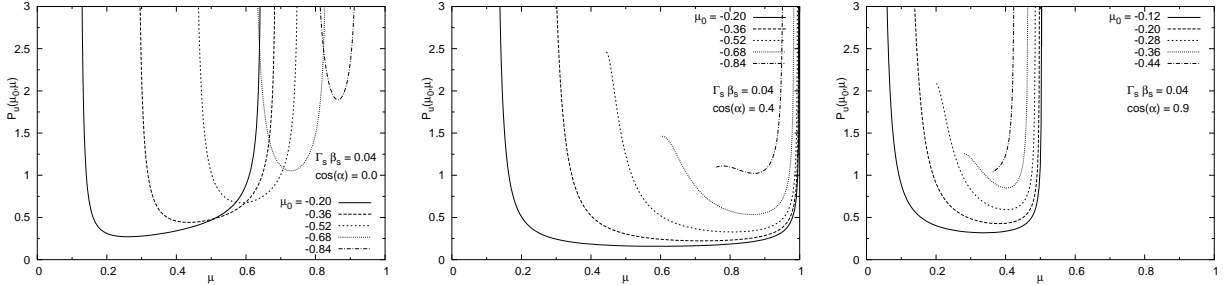}
\caption{Conditional probability $P_u(\mu_\circ,\mu)$ as a
function of the outgoing direction $\mu$, for a fixed value of the
shock speed ($\Gamma_s \beta_s= 0.04$) with three different inclinations
of the magnetic field ($\cos\alpha = 0.0, 0.4, 0.9$). For
each plot the different lines correspond to different values of
the ingoing direction $\mu_\circ$. \label{sez3fig_Pu_gb0.04}}
\end{figure}

\begin{figure}
\includegraphics[angle=0,scale=.62]{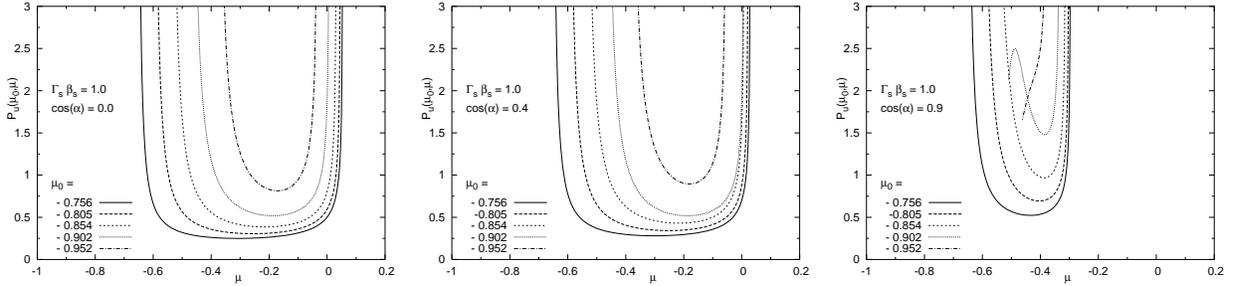}
\caption{Like fig.\ref{sez3fig_Pu_gb0.04} but for a trans-relativistic
shock ($\Gamma_s \beta_s= 1.0; \beta_s= 0.707$).
\label{sez3fig_Pu_gb1.0}}
\end{figure}

\begin{figure}
\includegraphics[angle=0,scale=.62]{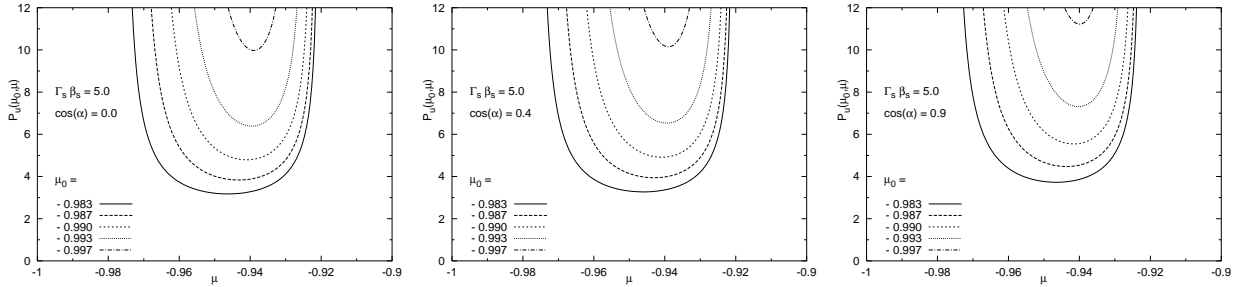}
\caption{Like fig.\ref{sez3fig_Pu_gb0.04} but for a relativistic shock
($\Gamma_s \beta_s= 5.0; \beta_s= 0.98$). The three plots are very
  similar to each other because the condition $\cos\alpha> \beta_s$
  is never reached.
\label{sez3fig_Pu_gb5.0}}
\end{figure}

Figs. \ref{sez3fig_Pu_gb0.04}, \ref{sez3fig_Pu_gb1.0} and
\ref{sez3fig_Pu_gb5.0} show some examples of our calculations of
$P_u(\mu_\circ,\mu)$ as a function of $\mu$ for different values
of $\mu_\circ$, for a Newtonian, a trans-relativistic and a
relativistic shock respectively. For each case we show the results for
different inclinations of the magnetic field with respect to the shock
normal.
It is worth noticing that $P_u$ does not change significantly when the
inclination of the magnetic field varies in the range
$0<\cos\alpha<\beta_s$, at a given shock speed. Therefore we do
not expect a significant variation of the spectral slope in this
range. In \S \ref{sec:sp_ani} we show that this is in fact the case.

\subsection{Spectrum and Anisotropy of the accelerated particles for a
large scale magnetic field upstream}\label{sec:sp_ani}

In this section we use Eq. (\ref{sez2eq_ang-dis}) and
Eq. (\ref{sez2eq_Q^T}) to calculate the spectrum and angular
distribution of the accelerated particles at the shock front.
The return probabilities are calculated assuming that in the
downstream fluid there is isotropic scattering, so that $P_d$ can be
calculated from Eq. (\ref{sez2eq_Pd}) using Eq. (\ref{sez2eq_w_iso})
as a scattering function. We assume $\sigma=0.01$ for the SPAS
regime and $\sigma=10$ for the LAS regime. In the upstream fluid
we assume that particles can only be deflected by a large scale
coherent magnetic field with arbitrary orientation with respect to the
normal to the shock front. The return probability $P_u$ is therefore
calculated as discussed in detail in \S \ref{sec:upP}.

The only information still lacking to proceed further is an equation
of state for the medium, that would allow us to compute the velocity
of the downstream fluid from the jump conditions at the shock front
(see for instance \cite{gal02}). We assume that the gas upstream has
zero pressure. Moreover, in the following we assume everywhere that
the magnetic field has no dynamical role, so that the standard jump
conditions for an unmagnetized shock can be adopted (the role of
the magnetic field becomes important when the magnetic energy density
becomes comparable with the thermal energy density \cite{kir99}).

Following much of the previous literature, we adopt the Synge
equation of state for the downstream gas \cite{syn57}, assuming that
only protons contribute. Although used widely, this assumption may not
be well justified in a general case. We will illustrate our
conclusions on the role of the equation of state for the spectrum and
anisotropy of the accelerated particles in a separate paper.

Within this set of assumptions it is worth reminding that the
compression ratio $u_{\rm up}/u_{\rm down}$ tends asymptotically to $4$
for a non relativistic shock (even for shock speeds that are known to
give lower compression factors) and to $3$ for ultra-relativistic
shocks.

\begin{figure}
\begin{center}
\plottwo{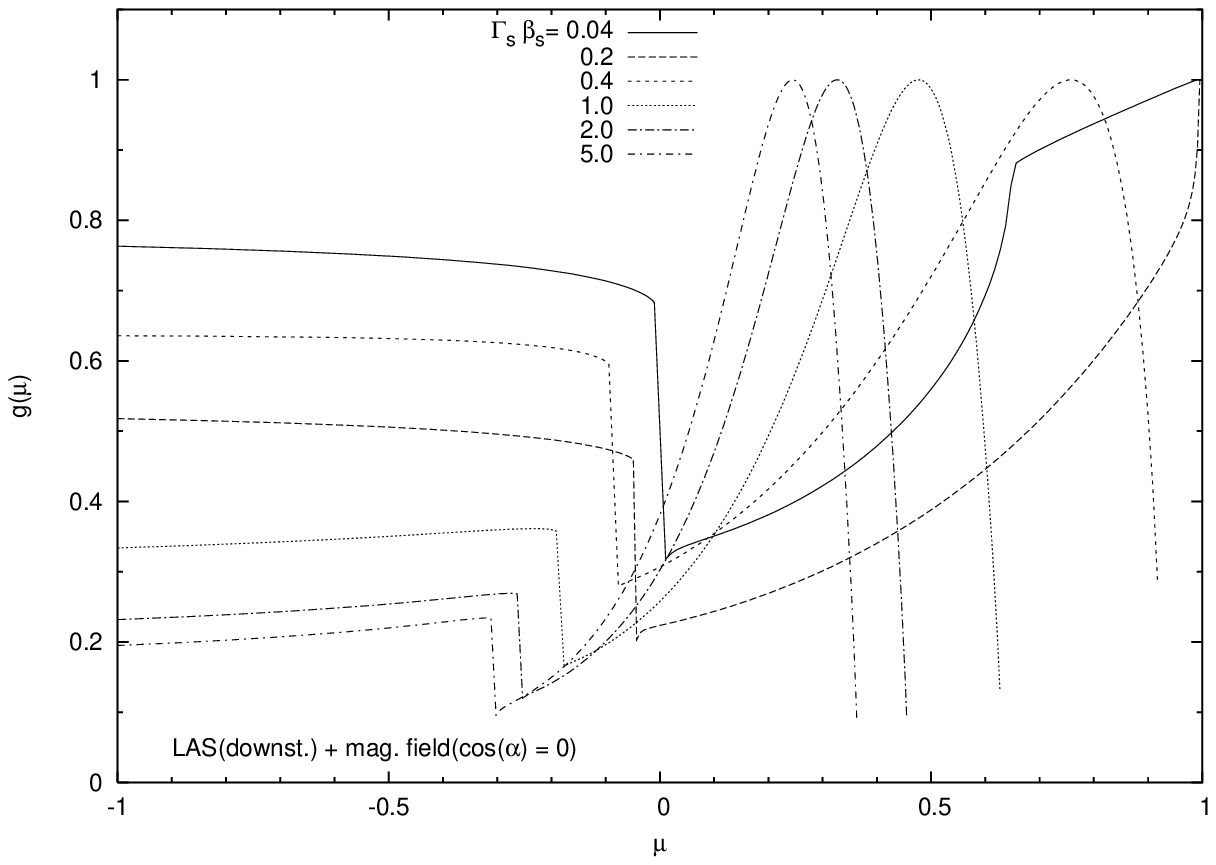} {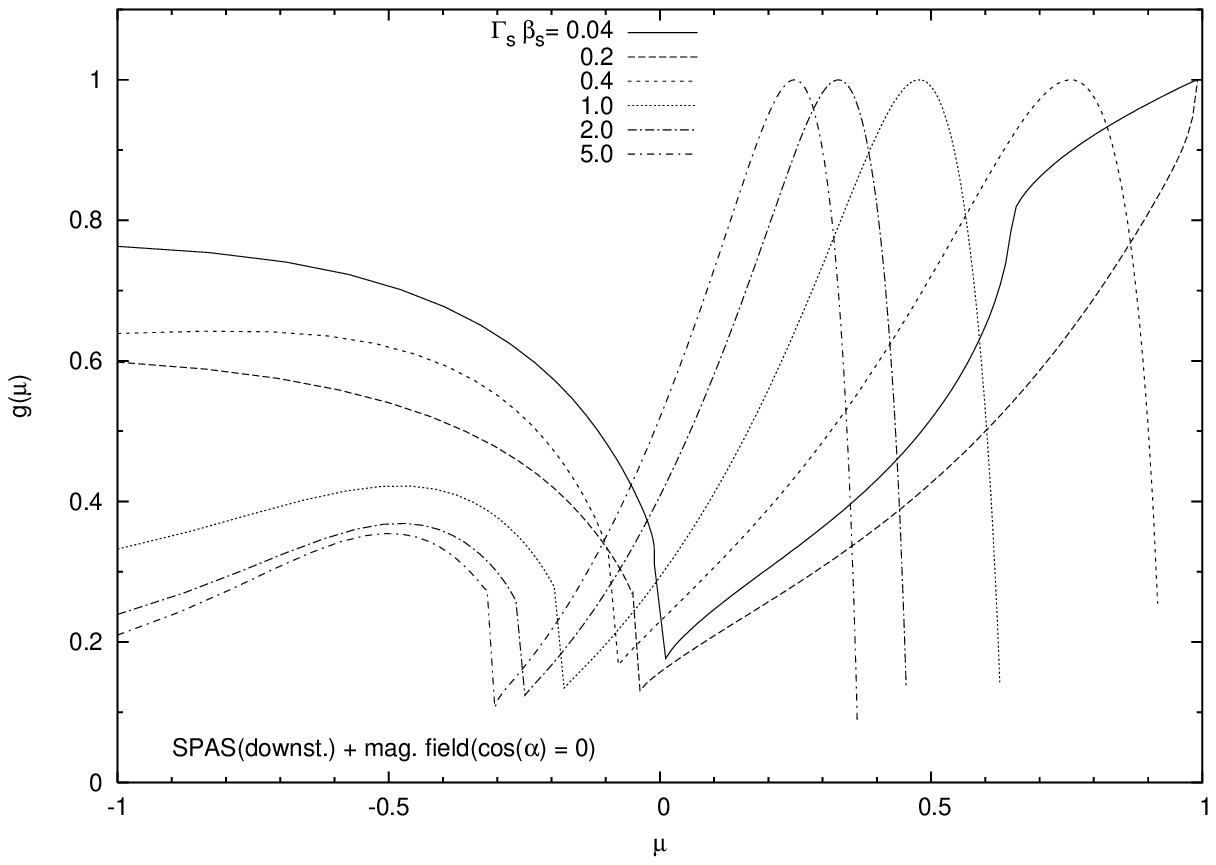} \caption{Particle distribution function
at the shock front when a large scale coherent magnetic field is
present in the upstream region, with a direction parallel to the
shock plane. In the downstream region particles are scattered in the
LAS (left plot) or in the SPAS regime (right plot) (here the maximum
of is arbitrarily set equal to 1). Several values of shock speeds
are shown. The particle distribution functions always show a jump at
$\mu=-\beta_s$. Large angle scattering makes distribution functions
flatter compared with the small angle scattering case for
$-1<\mu<-u_d$. \label{sez3fig_ang_dis_B}}
\end{center}
\end{figure}

\begin{figure}
\begin{center}
\includegraphics[angle=0,scale=1]{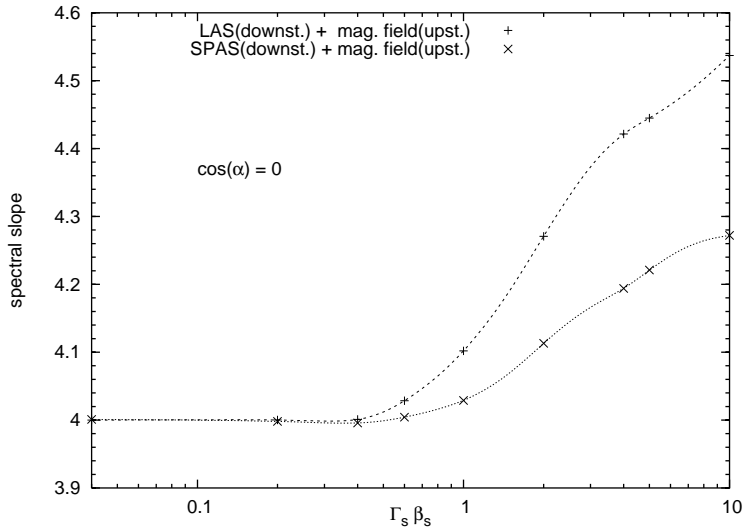}
\caption{Spectral index vs. shock speed for the same configuration as in
fig.\ref{sez3fig_ang_dis_B}.
\label{sez3fig_slope_LAS+B-SPAS+B}}
\end{center}
\end{figure}

The simplest case to consider is that of a shock in which the large
scale coherent magnetic field in the upstream region is parallel to
the shock front ($\cos\alpha=0$). This is known as a perpendicular shock.
The angular distribution and the slope of the spectrum of the
accelerated particles are plotted in Fig. \ref{sez3fig_ang_dis_B}
(the LAS (SPAS) case is shown in the left (right) panel) and Fig.
\ref{sez3fig_slope_LAS+B-SPAS+B} respectively, for various shock
velocities ranging from newtonian to relativistic.

The angular distribution of the particles in the downstream frame is
seen to be rather anisotropic for the SPAS case, even in the newtonian
regime. Large angle scattering (LAS) is evidently more efficient in
isotropizing the accelerated particles. The anisotropies do
not seem to affect the spectrum of the accelerated particles in the
case of non relativistic shocks: the slope of the spectrum for both
SPAS and LAS is $4.000\pm0.001$. The effect becomes more prominent for
faster shocks and in particular for relativistic shocks.
In the SPAS case, for $\Gamma_s \beta_s = 10$, we found
$s=4.272\pm0.001$, compatible with  $s = 4.28\pm 0.01$,  obtained
by \cite{ach01} for $\Gamma_s = 10$, with a Monte-Carlo simulation.

In Fig. \ref{sez3fig_slope_LAS+B-SPAS+B}, the dotted and dashed lines
refer to the SPAS and LAS cases respectively. At first sight it may
appear rather surprising that in the limit of relativistic shocks the
spectrum of accelerated particles is softer in the LAS regime than it
is in the SPAS regime, since LAS is envisioned as more efficient in
redirecting the particles to the shock front. This intuitive vision
turns out to be incorrect, as also shown in Table 1, where we list
the slope, the average energy gain and the return probability from
downstream (as defined in Eqs. (\ref{sez3eq_Pd_mean}) and
(\ref{sez3eq_mean_amp})) for a relativistic shock with $\Gamma_s \beta_s = 5.0$.

\begin{table}
\begin{center}
\caption{Exact spectral slope, mean amplification and downstream
return probability (as defined in eq. (\ref{sez3eq_Pd_mean}) and
(\ref{sez3eq_mean_amp}) respectively) for $\Gamma_s \beta_s = 5.0$.
\label{sez3tab_SPAS-LAS}}
\begin{tabular}{crrrr}
\tableline\tableline
 & slope & $\langle G \rangle$ & $\langle P_{\rm ret}^{\rm (down)} \rangle$ & \\
\tableline
SPAS &$4.218\pm0.001$ &2.0387 &0.4165 \\
LAS  &$4.445\pm0.001$ &2.0753 &0.3430 \\
\tableline
\end{tabular}
\end{center}
\end{table}

\begin{figure}
\begin{center}
\includegraphics[angle=0,scale=1]{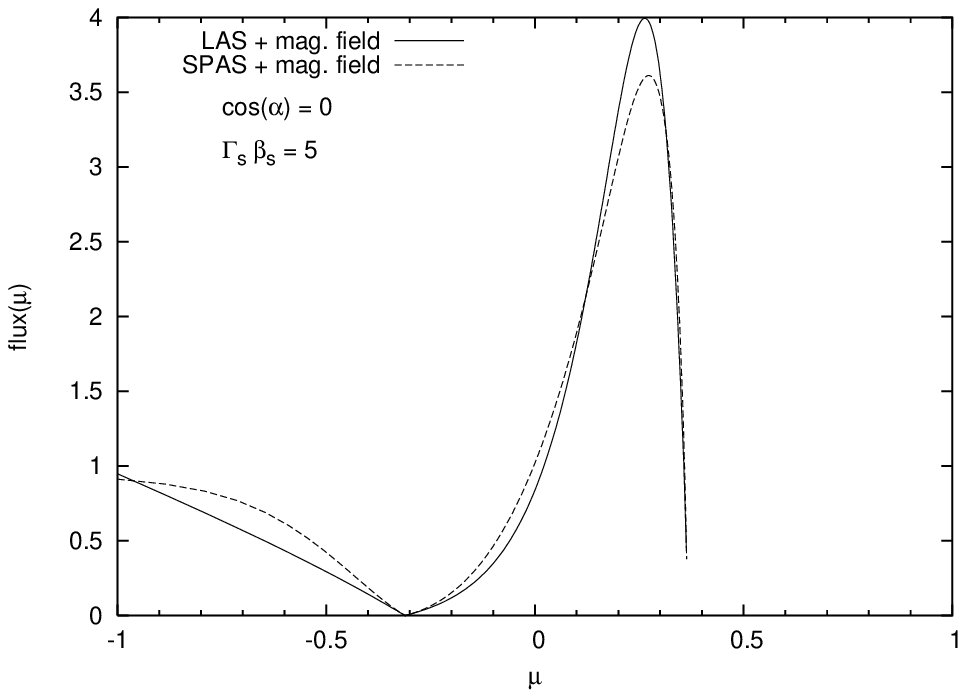}
\caption{Particle flux $J(\mu) \equiv |\mu+u_d| g(\mu)$ across the
  shock front, as it appears in the downstream frame, when the upstream
magnetic field is parallel to the shock ($\cos\alpha= 0$) and the
downstream fluid is in the LAS regime (solid line) or SPAS regime
(dashed line). The shock speed is $\Gamma_s \beta_s = 5.0$.
The flux entering downstream (\textit{i.e.}  for $-u_d<\mu<1$) is
normalised to $1$. In this way we see that downstream return
probability, \textit{i.e.}  the integrated flux for $-1<\mu<-u_d$,
is larger when the downstream region is in the SPAS regime.
\label{sez3fig_flux_LAS+B-SPAS+B_gb5}}
\end{center}
\end{figure}

One can see that while the average energy gain is similar in the two
cases, the return probability in the case of LAS is 20\% lower than
for the SPAS case. Qualitatively this can be understood as follows:
when the shock velocity increases, particles are caught up by the
shock front when they have travelled only a small fraction (of order
$1/\Gamma_{s}$) of their gyration. Once downstream, LAS is likely to
swing them far from the shock front in a few interactions, while SPAS
deflects their trajectories rather slowly yet remaining in the
vicinity of the shock surface. This is responsible for the 20\%
difference in the average return probabilities in the two cases. This
is also shown in Fig. \ref{sez3fig_flux_LAS+B-SPAS+B_gb5}, where we
plot the particles flux, $J(\mu) \equiv |\mu+u_d| g(\mu)$ in terms of
downstream coordinates: the total flux of particles entering the
downstream section ($-u_d<\mu<1$) is normalized to unity. It is clear
from Fig. \ref{sez3fig_flux_LAS+B-SPAS+B_gb5} that the flux of
particles returning to the shock is slightly larger for the case of
SPAS (dashed line in the range $-1<\mu<-u_d$).

A more interesting question concerns the effect of the orientation
of the large scale magnetic field with respect to the normal to the
shock. We have already emphasized that for any orientation different
from that of a perpendicular shock, and in the absence of scattering
processes upstream, particles are lost from the upstream region,
because the shock cannot catch up with their motion. This happens
when $\cos\alpha>-\beta_s$, so that the phenomenon is increasingly
more important for shocks approaching the parallel configuration.
This reflects in increasingly softer spectra. In the limit
$\cos\alpha \rightarrow 1$, all particles escape from the upstream
region and no acceleration takes place.

\begin{figure}
\begin{center}
\includegraphics[angle=0,scale=1]{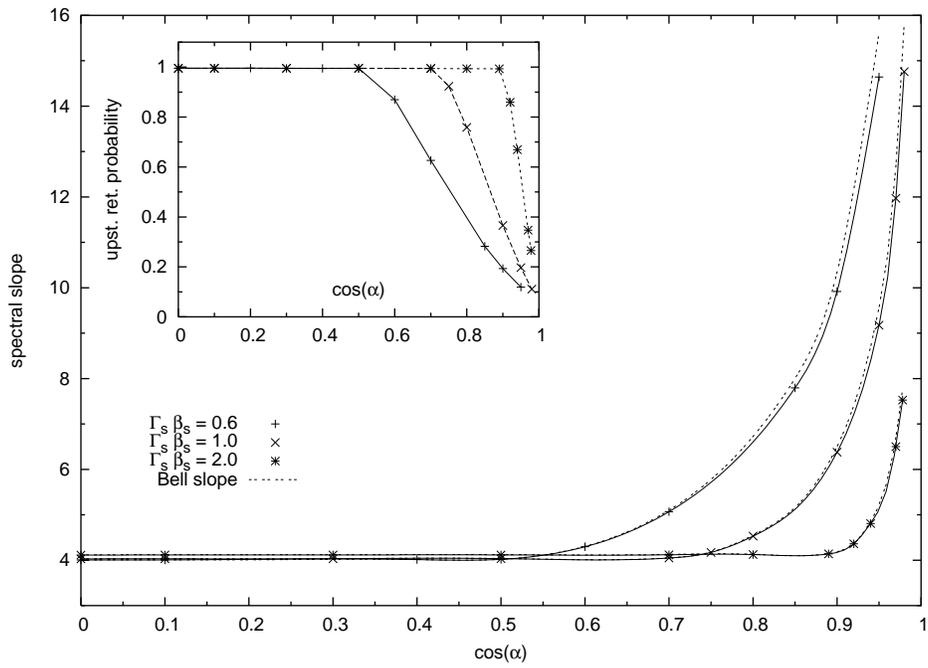}
\caption{Large box: spectral slope as a function of $\cos\alpha$ for three
different values of the shock speed. The dashed lines show the spectral
slope computed with Bell's method.
Small box: the corresponding upstream return probabilities.
\label{sez3fig_slope_cos(alpha)}}
\end{center}
\end{figure}

The slope of the spectrum as obtained from our calculations is
plotted in Fig. \ref{sez3fig_slope_cos(alpha)} (solid lines and
symbols) as a function of $\cos\alpha$ for three different shock
speeds ($\Gamma_s \beta_s = 0.6, 1.0, 2.0$): when there is no
particle escape, the slope $s$ is actually a constant, while it
increases dramatically (and in fact diverges, showing the
disappearance of the acceleration process) for values of
$\cos\alpha$ larger than $-\beta_s$. In the small panel in Fig.
\ref{sez3fig_slope_cos(alpha)} we also plot the return probability
from upstream: for very inclined shocks the return probability is
still very close to unity, as in the case of upstream scattering,
but it drops rapidly for increasingly less inclined shocks.

The steepening of the spectrum due to leakage of the particles
towards upstream infinity can also be understood in terms of a
Bell-like \cite{bel78} calculation, when carried out for the case of
a large scale coherent magnetic field. The slope of the spectrum is
related to the average return probability and to the average energy
gain of the particles per cycle back and forth through the shock
front through the expression:

\begin{equation} \label{sez3eq_Bell_slope}
s = 3- {\log \langle P_{\rm ret}\rangle \over \log \langle G \rangle } ,
\end{equation}
where $\langle G \rangle$ is the mean amplification in a single
cycle (downstream $\rightarrow$ upstream $\rightarrow$ downstream),
and $\langle P_{\rm ret}\rangle$ is the mean probability of
returning to the shock. One should keep in mind that Bell's method,
as expressed through the equation above is flawed in that it does
not take into proper consideration the correlation between the
amplification factor and the return probability. Moreover, Eq.
(\ref{sez3eq_Bell_slope}) hides the assumption of isotropy of the
distribution function of the accelerated particles, since that
formula was conceived in a discussion of non relativistic shocks
(\cite{pea81} introduced this formalism for particle acceleration at
relativistic shock fronts). All these limitations become of
particular importance for relativistic shocks. A general expression
for the slope was found in \cite{vie03}, and reads:
\begin{equation} \label{sez3eq_Bell_slope_corr}
\langle P_{\rm ret}\rangle \langle G^{s-3} \rangle = 1.
\end{equation}
In the following we use Eq. (\ref{sez3eq_Bell_slope}), since we only
want to provide the reader with an argument of plausibility for the
steepening of the spectra in those cases in which particle leakage
can take place in the upstream region. In order to account for this
leakage, which cannot take place in the standard scenario of
diffusive particle acceleration at a shock front, we generalize Eq.
(\ref{sez3eq_Bell_slope}) in order to include the probability of
escape from the acceleration box from upstream. This is easily
achieved by replacing  $\langle P_{\rm ret} \rangle$ with $\langle
P_{\rm ret}^{\rm (up)} \rangle \cdot \langle P_{\rm ret}^{\rm
(down)} \rangle$. These mean values expressed in the downstream
frame are:
\begin{equation} \label{sez3eq_Pd_mean}
\langle P_{\rm ret}^{\rm (down)}\rangle =
\frac{\int_{-1}^{-u_d}d\mu_\circ \,
\int_{-u_d}^{1}d\mu\, g(\mu) (u_d+\mu)
P_d(\mu,\mu_\circ)}{\int_{-u_d}^{1} d\mu \, g(\mu) (u_d+\mu)}
\end{equation}
and
\begin{eqnarray} \label{sez3eq_Pu_mean}
\langle P_{\rm ret}^{\rm (up)}\rangle = \frac{\int_{-u_d}^{1}d\mu
\int_{-1}^{-u_d}d\mu_\circ \, g(\mu_\circ) (u_d+\mu_\circ)
P_u(\mu_\circ,\mu)}{\int_{-1}^{-u_d} d\mu_\circ \, g(\mu_\circ) (u_d+\mu_\circ)}
\, .
\end{eqnarray}
In the last equation $P_u$ has also to be computed in terms of
quantities evaluated in the downstream frame. Energy amplification
for a particle entering the upstream region with direction $\mu_\circ$
(as measured downstream) and returning with direction $\mu$, is
obtained combining two Lorentz transformations:
\begin{equation} \label{sez3eq_amp}
G(\mu_\circ,\mu) = \gamma_{\rm rel}^2 (1- u_{\rm rel}\mu_\circ) \,
\left( 1+ u_{\rm rel} \bar \mu \right) \, ,
\end{equation}
where $\bar\mu = (\mu+u_{\rm rel}) / (1+u_{\rm rel} \mu)$ is the
returning direction as seen in the upstream frame.
Averaging the amplification  we have:
\begin{equation} \label{sez3eq_mean_amp}
\langle G \rangle = \frac{\int_{-1}^{-u}d\mu_\circ \,
g(\mu_\circ) (u+\mu_\circ) \int_{-u}^{1} d\mu \,
G(\mu_\circ,\mu) P_u(\mu_\circ,\mu)}{ \int_{-1}^{-u} d\mu_\circ \,
g(\mu_\circ) (u+\mu_\circ) \int_{-u}^{1} d\mu \, P_u(\mu_\circ,\mu)} \, .
\end{equation}
The spectral slope as computed through Eq. (\ref{sez3eq_Bell_slope})
is plotted in Fig. \ref{sez3fig_slope_cos(alpha)} (large box) with
dashed lines; the corresponding upstream return probability
$\langle P_{\rm ret}^{\rm (up)}\rangle$ is plotted in the small box
(dashed lines). The agreement with our exact results is better than
$1\%$, proving that the reason for the softening of the spectra of
accelerated particles is in the increased probability that the
particles leave the acceleration region when only a large scale
coherent magnetic field is present upstream.

The results discussed above apply to situations in which the
magnetic field in the upstream region can be considered as coherent
on spatial scales exceeding the size of the acceleration box. If the
coherence scale of the field is smaller than the size of the
accelerator, then the direction of the particles suffer a random
wandering motion and one can think of this structured field as the
source of diffusion and as a physical mechanism that imposes a
maximum energy to the accelerated particles (at least in the absence
of radiative energy losses). Particles that escape from the shock
region too fast (highest energy ones) have enough time to {\it feel}
the effect of a coherent scale, while lower energy particles live in
the accelerator for longer times and in principle may {\it feel}
different orientations of the upstream magnetic field. This scenario
is basically equivalent to having some degree of scattering
upstream, and should be treated with the formalism already discussed
in \cite{vie03,bla05}. As soon as a phenomenon equivalent to
scattering is present, the probability of escape to upstream
infinity vanishes, for all those particles that are confined in the
accelerator for sufficiently long times. Moreover, one should keep
in mind that even if a large scale coherent magnetic field is
present to start with, the propagation of the accelerated (charged)
particles in the upstream plasma is very likely to excite
fluctuations in the magnetic field structure through streaming
instability \cite{bel78}. These fluctuations act as scattering
centers and enhance the probability of returning to the shock front.

\section{Anisotropic scattering}\label{sec:ani}

In this section we consider again the standard case in which particle
motion in both the upstream and downstream fluids is diffusive, due to
the presence of scattering agents. However, we include the possibility
that the scattering, though spatially constant, may be anisotropic.
The physical motivation for this generalization is the following:
in a background of Alfv\`en waves with a power spectrum $P_W(k)$
(such that $P_W(k)dk$ is the energy density in the form of
waves with wavenumber in the range $dk$ around $k$) the particles
suffer angular diffusion with a diffusion coefficient
\begin{equation}
D_{\theta\theta} = \langle \frac{\Delta \theta \Delta \theta}{\Delta
  t} \rangle \approx \Omega \frac{k_r P_W(k_r)}{B_0^2/8\pi},
\label{eq:res}
\end{equation}
where $k_r = \Omega/v \mu$ is the resonant wavenumber and $\Omega$ is
the gyration frequency of particles with momentum $p$ in the
background magnetic field $B_0$. One can clearly see from
Eq. (\ref{eq:res}) that the diffusion is anisotropic in general, unless
the power spectrum has a specific {\it ad hoc} form. One should keep
in mind that Eq. (\ref{eq:res}) is obtained in the context of
quasi-linear theory. A full non-linear treatment might show how the
turbulence is distributed and which is the resulting particle
angular distribution.

In the calculations that follow, we quantify the effects of anisotropic
scattering on the spectrum and angular distribution of the accelerated
particles. The calculation of specific patterns of anisotropy in the
scattering agents is beyond the scopes of this paper, therefore we
adopt a few simple but physically meaningful toy models of anisotropic
scattering and we carry out the calculations within those models.

\subsection{Modelling anisotropy}\label{sec:model_ani}

We parametrize the anisotropy in such a way to reproduce the following
four patterns:

\begin{itemize}
\item \textit{case A}: Particles are scattered per unit length more
  efficiently while they move away from the shock front than they are
  on their way to the shock front, both upstream and downstream.

\item \textit{case B} (opposite of case \textit{A}): Particles are
  scattered per unit length more efficiently on their way to the shock
  front than they are while they move away from the shock front, both
  upstream and downstream.

\item \textit{case C}: In the downstream fluid, particles are
  scattered per unit length more efficiently while they move away from
  the shock front ($\mu\to 1$) than they are on their way to the shock
  front ($\mu\to -1$). In the upstream fluid the situation is
  reversed, and scattering is more efficient for the particles that
  are moving toward of the shock ($\mu\to 1$).

\item \textit{case D} (opposite of \textit{C}): Scattering is more
  effective around $\mu\sim -1$ both upstream and downstream.

\end{itemize}

\begin{figure}
\begin{center}
\plottwo{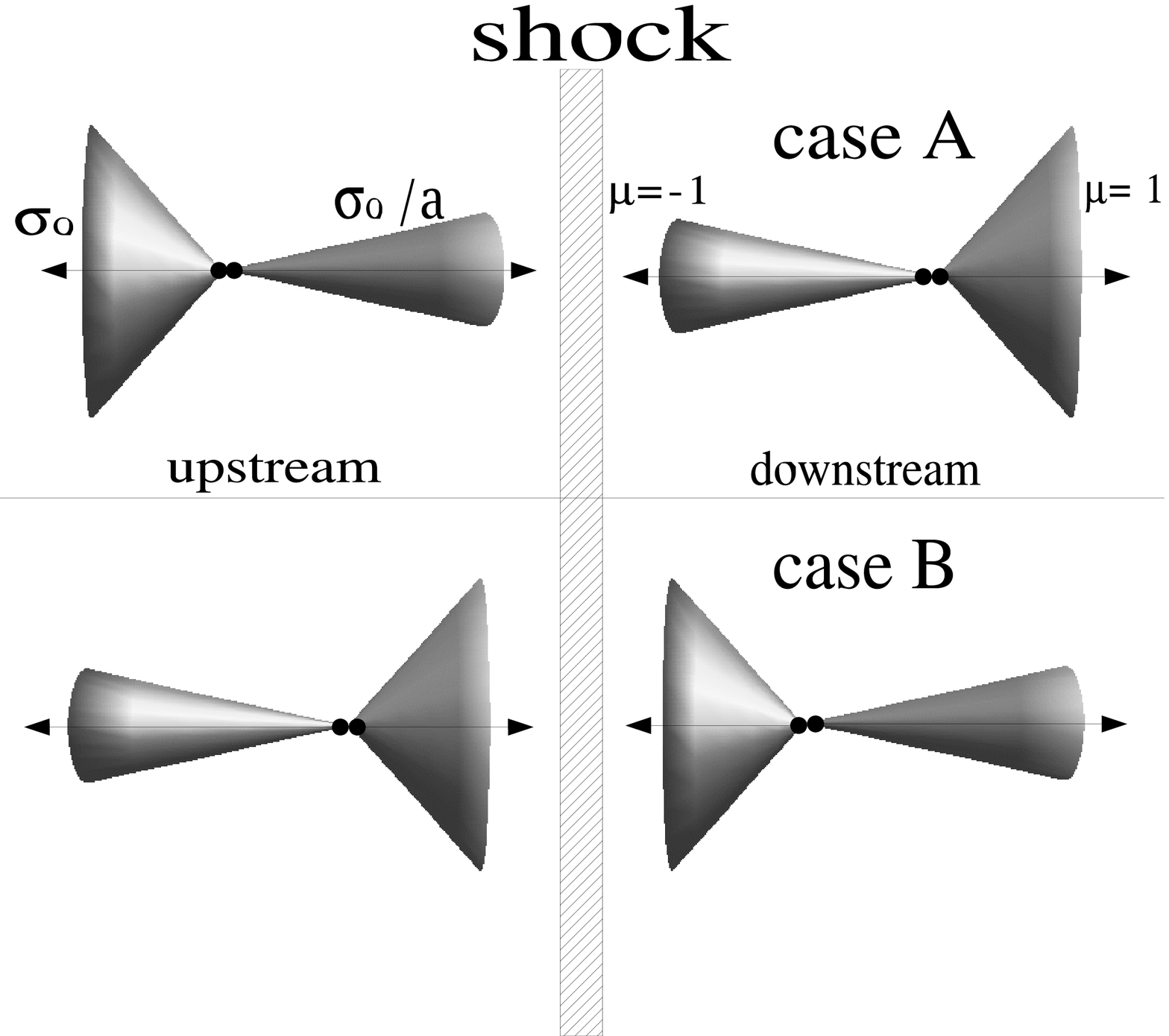}{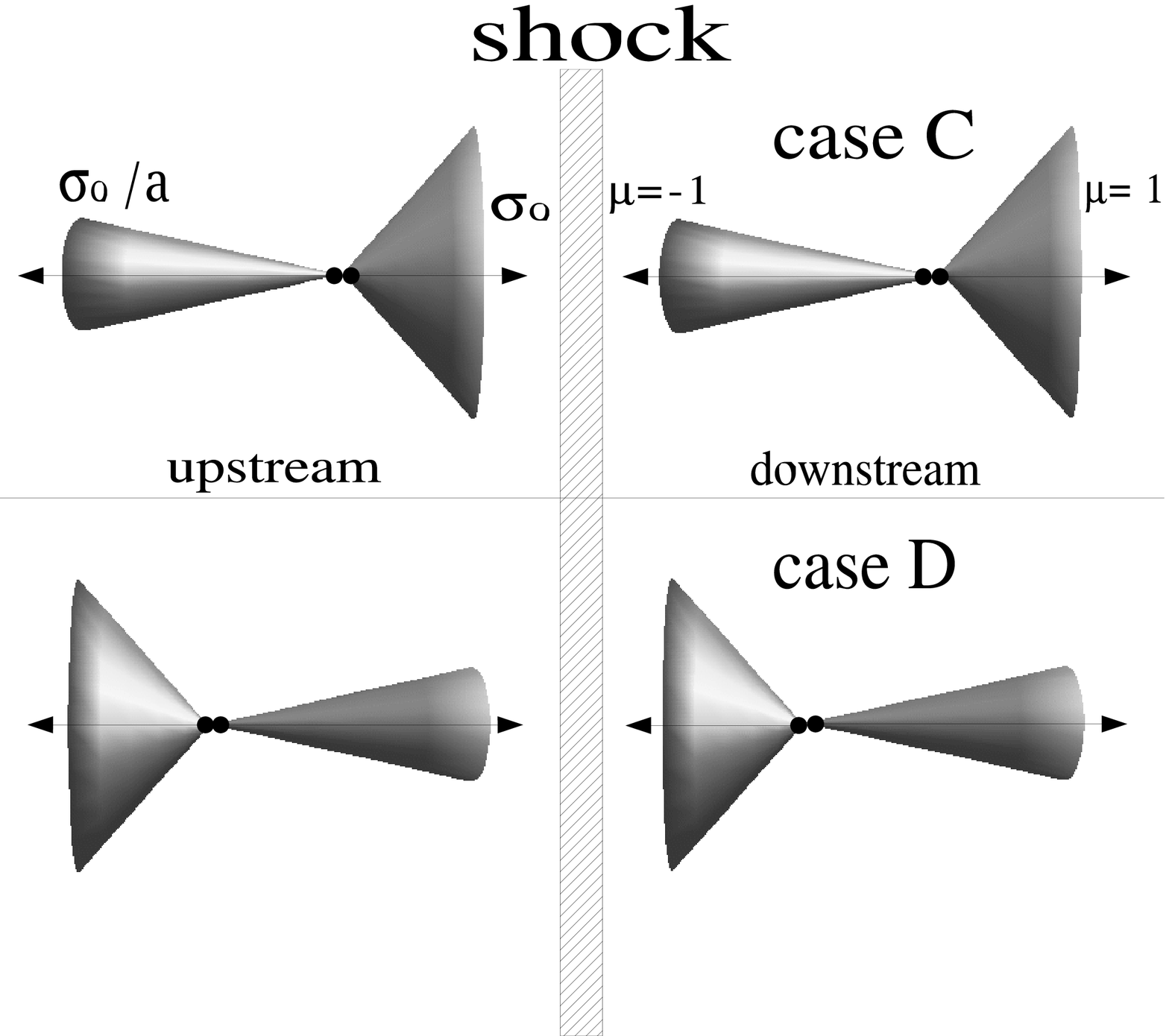} \caption{Pictorial representation of the
four patterns of anisotropic
  scattering considered in our calculations.
\label{sez3fig_anis_skeme}}
\end{center}
\end{figure}

A pictorial representation of cases \textit{A-D} is shown in Fig.
\ref{sez3fig_anis_skeme}.

In order to simulate the cases A-D above, we adopt a scattering
function similar to Eq. (\ref{sez2eq_w_iso}), but modified to introduce
anisotropic scattering. In particular, to achieve this goal we allow
the width $\sigma$ of the scattering function to depend on both the initial and
final directions $\mu'$ and $\mu$, so that:

\begin{equation} \label{sez4eq_w_anis}
w(\mu,\mu') = {1\over \sigma(\mu,\mu')} e^{-{1-\mu\mu' \over
    \sigma(\mu,\mu')}}
I_0\left( {\sqrt{1-\mu^2} \sqrt{1-\mu'^2} \over \sigma(\mu,\mu')}\right) \, .
\end{equation}

It is worth stressing that the scattering function has to be symmetric 
if we exchange $\mu$ with $\mu'$ as a consequence of Liouville's theorem, so we
are forced to look for a symmetric function $\sigma(\mu,\mu')$.

In order to apply the functional form Eq. (\ref{sez4eq_w_anis}) to the cases
A-D, it is sufficient to adopt the following expression for the mean 
scattering angle $\sigma(\mu,\mu')$:
\begin{equation} \label{sez4eq_sigma_anis}
\sigma_\mp(\mu,\mu') = \sigma_\circ \cdot \left( 1- {(a-1) \over 4 a}
(\mu \mp 1)(\mu' \mp 1) \right) \, .
\end{equation}
Both $\sigma_+$ and $\sigma_-$  have  $\sigma_0$ as the maximum and
$\sigma_\circ/a$ as the minimum value. For this reason we will refer to $a$ as
the {\it Anisotropy Factor}. For $a=1$, isotropic scattering is recovered.

\begin{figure}
\begin{center}
\plottwo{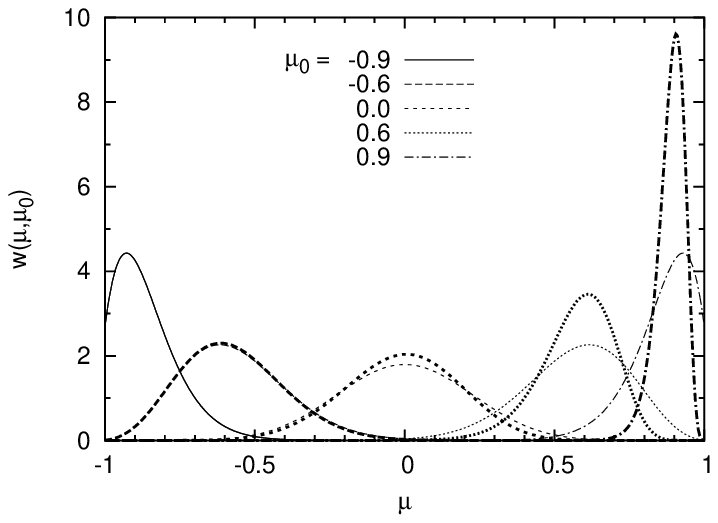}{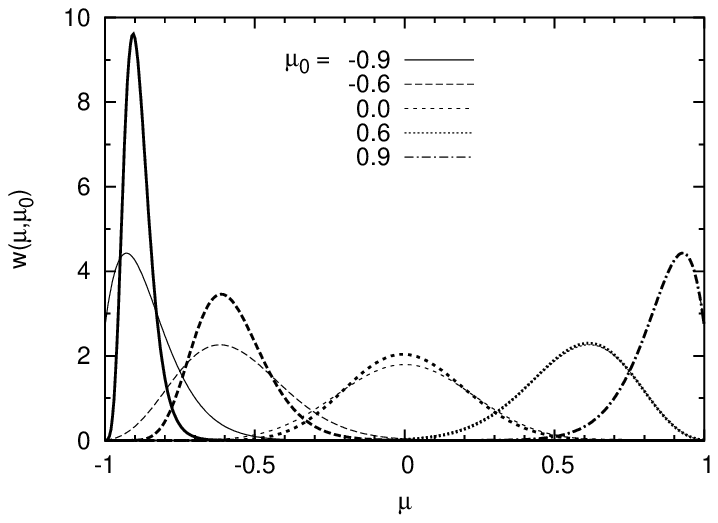} \caption{The thick lines
show the anisotropic scattering functions $w_+(\mu,\mu_\circ)$ (left box) and
$w_-(\mu,\mu_\circ)$ (right box), as functions of $\mu$ and for different values
of the incoming direction $\mu_\circ$. The anisotropy factor is $a=10$ and
$\sigma_\circ=0.05$. 
For comparison the isotropic scattering function (Eq. (\ref{sez2eq_w_iso})) is
shown with thin lines and for the same values of $\mu_\circ$.
\label{sez4fig_scat-anis-func}}
\end{center}
\end{figure}

The resulting scattering function $w_\mp(\mu,\mu')$, obtained
substituting Eq. (\ref{sez4eq_sigma_anis}) into (\ref{sez4eq_w_anis}), is
plotted in Fig. \ref{sez4fig_scat-anis-func} together with the isotropic
scattering function (Eq.
(\ref{sez2eq_w_iso})), for $\sigma_\circ=0.05$ and $a=10$. 
These plots clarify how $w_+$ and $w_-$ can simulate a scattering more
efficient in the $\mu=+1$ and $\mu=-1$ directions respectively.

The condition $\int_{-1}^{1} d\mu \, w(\mu,\mu') = 1$ that states the
probability conservation, is fulfilled by Eq. (\ref{sez4eq_w_anis}) provided
$\sigma_0\ll 1$. In the numerical calculations that follow we assume
$\sigma_0=0.05$.

Using $\sigma_-$ and $\sigma_+$ in different combinations for the
upstream and the downstream fluids, we can reproduce scenarios
\textit{A}, \textit{B}, \textit{C}, and \textit{D}, as
summarized in Table \ref{sez4tab_ABCD}.
\begin{table}
\begin{center}
\caption{Summary of mean scattering angle used in the different
scenario of fig.\ref{sez3fig_anis_skeme}.\label{sez4tab_ABCD}}
\begin{tabular}{ccrrrr}
\tableline\tableline
 & $A$ & $B$ & $C$ & $D$ &\\
\tableline
upstream    &$\sigma_+$ &$\sigma_-$ &$\sigma_-$ &$\sigma_+$ \\
downstream  &$\sigma_-$ &$\sigma_+$ &$\sigma_-$ &$\sigma_+$ \\
\tableline
\end{tabular}
\end{center}
\end{table}

\subsection{Results: anisotropic scattering for shocks of arbitrary speed and
fixed anisotropy factor}
\label{sec:ani_rel}

Following the procedure outlined in \S \ref{sec:exact} and making use of
Eqs. (\ref{sez4eq_w_anis}) and (\ref{sez4eq_sigma_anis}), we compute the
spectral index and the angular distribution for the scenarios \textit{A},
\textit{B}, \textit{C} and \textit{D}, described above. 
In each case both the parameter $\sigma_\circ$ and the anisotropy factor $a$ are
fixed ($\sigma_\circ=0.05$ and $a=10$), while the shock velocity is allowed to
vary  within the range $0.04\leqslant \Gamma_s \beta_s \leqslant 5$.

\begin{figure}
\begin{center}
\plottwo{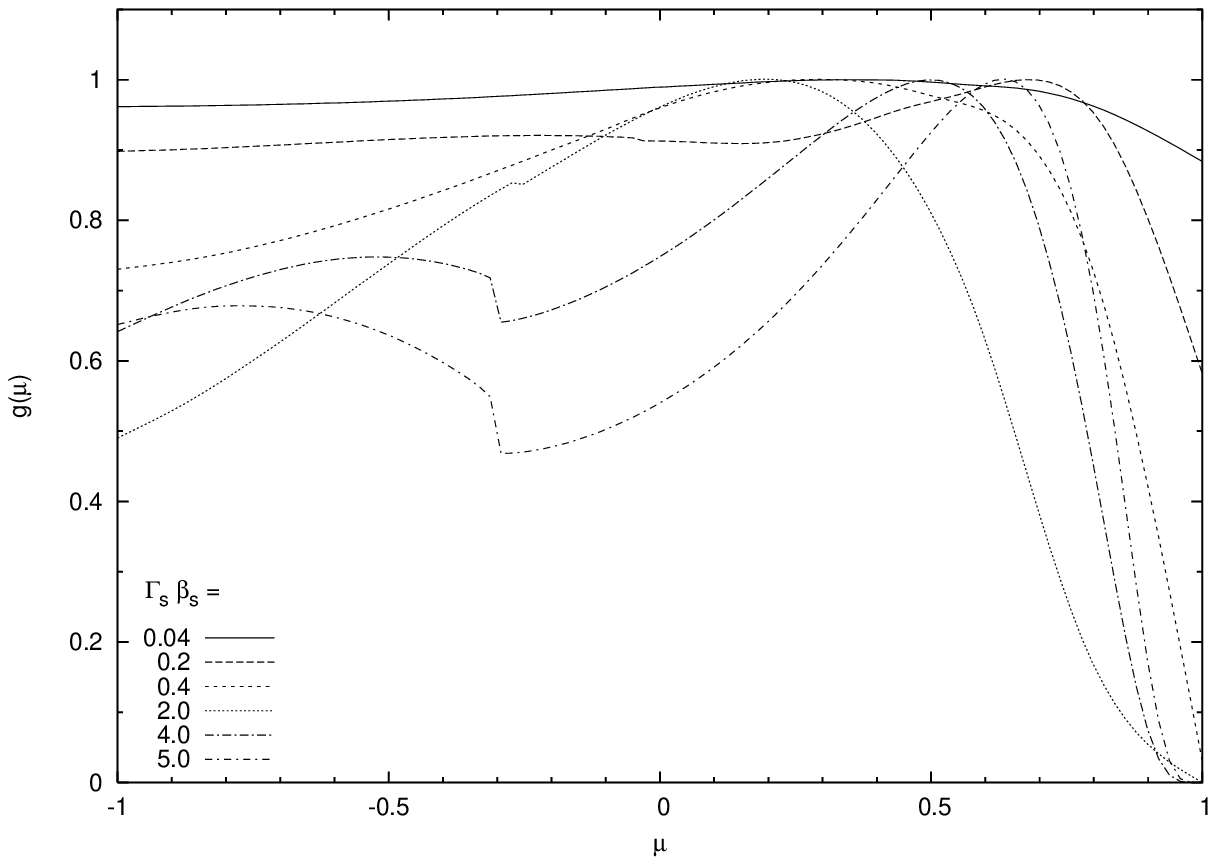} {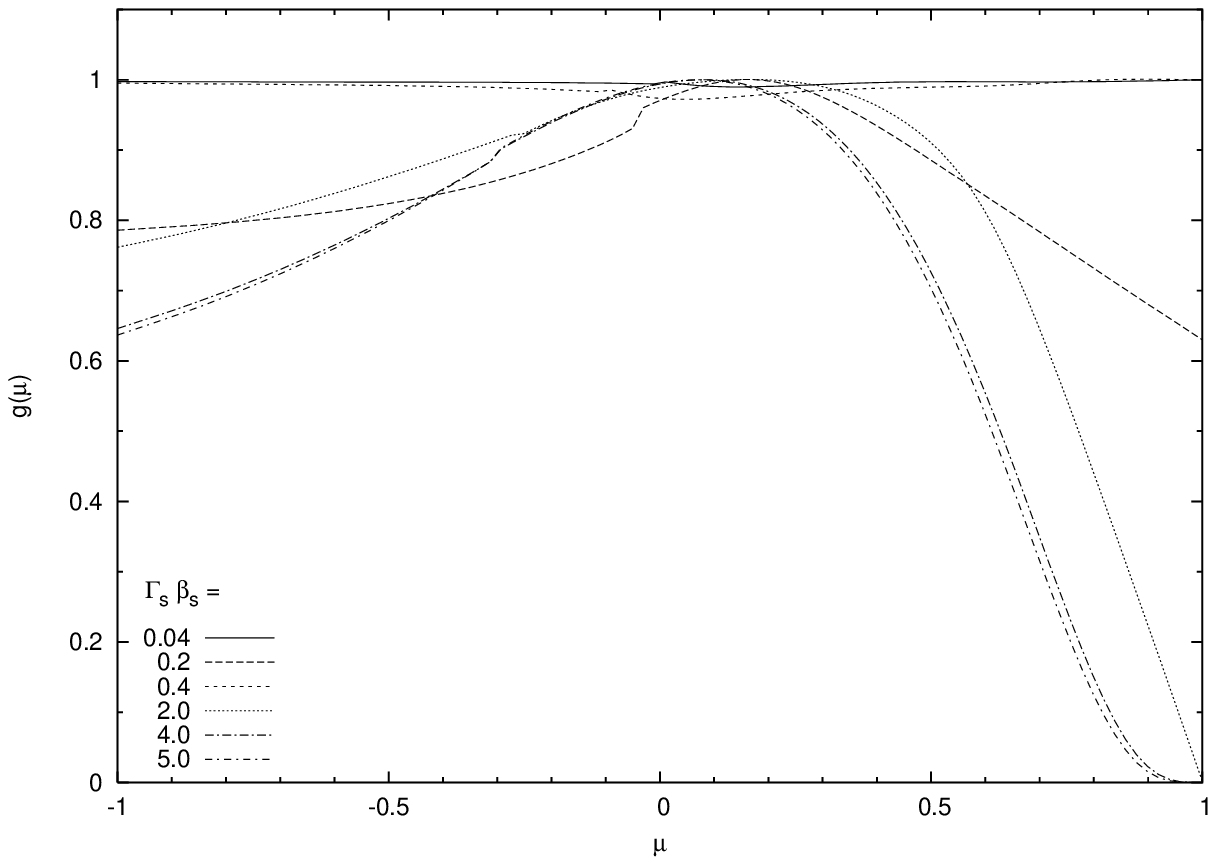} \caption{Particle distribution
function at the shock front for anisotropic scattering of type
\textit{A} (on the left) and \textit{B} (on the right) both with
$a=10$. Each line represents a different shock speed as the labels
show. \label{sez3fig_ang_dis_anis_AB}}
\end{center}
\end{figure}

\begin{figure}
\begin{center}
\plottwo{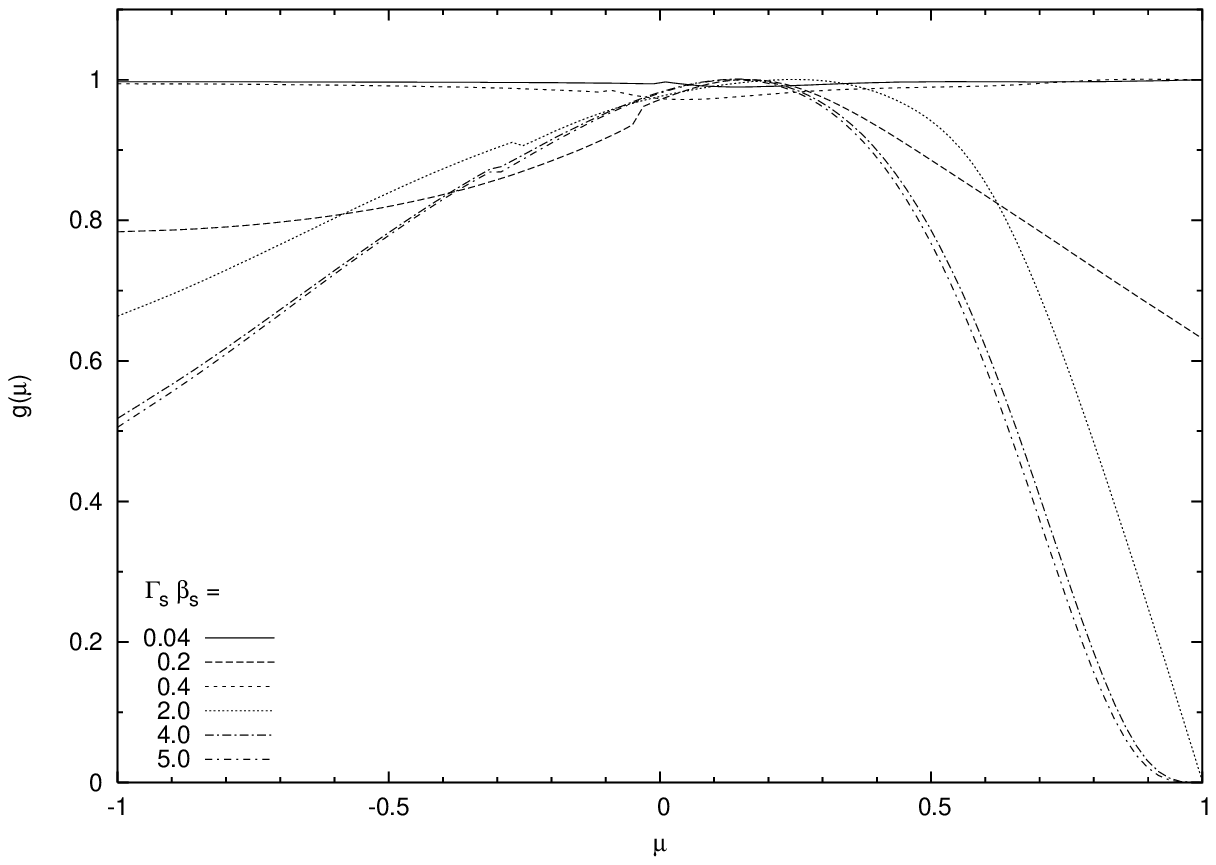} {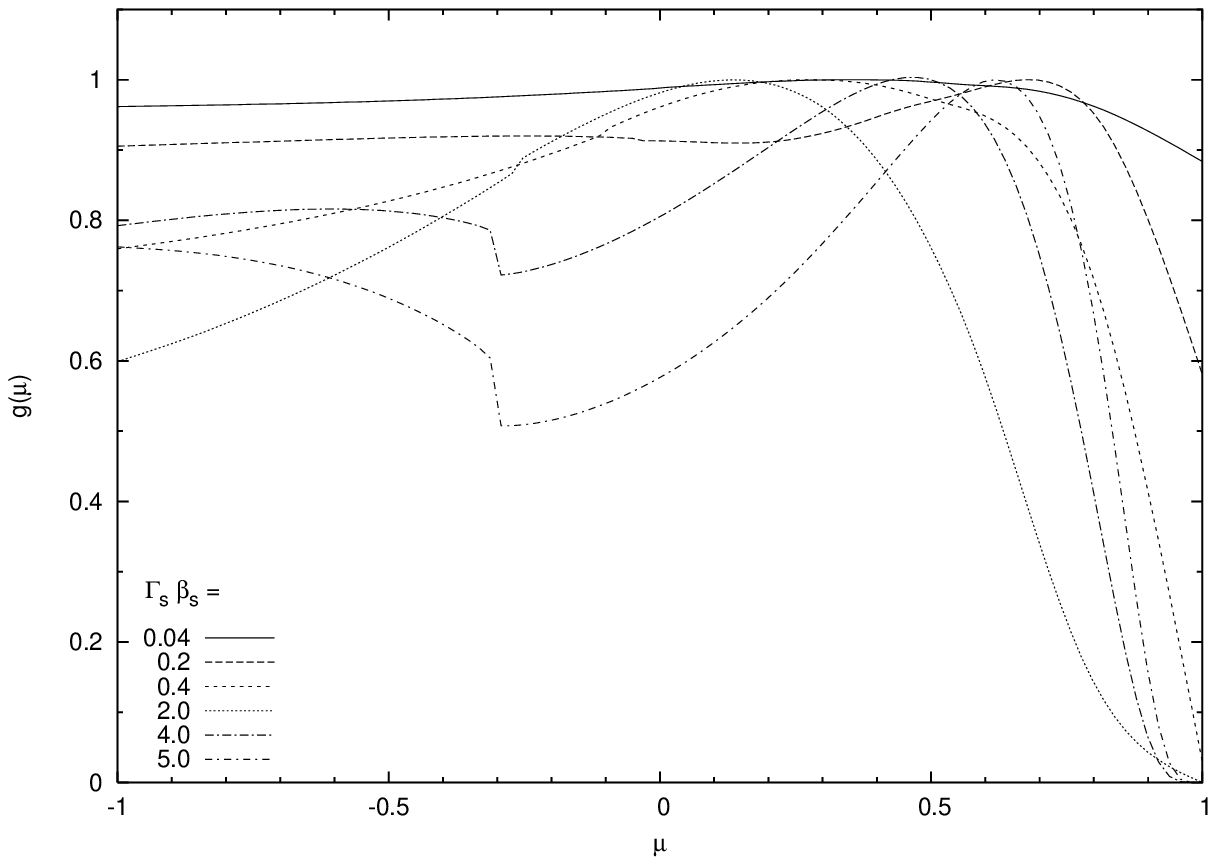} \caption{The same as
fig.\ref{sez3fig_ang_dis_anis_AB} but for anisotropic scattering of
type \textit{C} (on the left) and \textit{D} (on the right).
\label{sez3fig_ang_dis_anis_CD}}
\end{center}
\end{figure}

\begin{figure}
\begin{center}
\plottwo{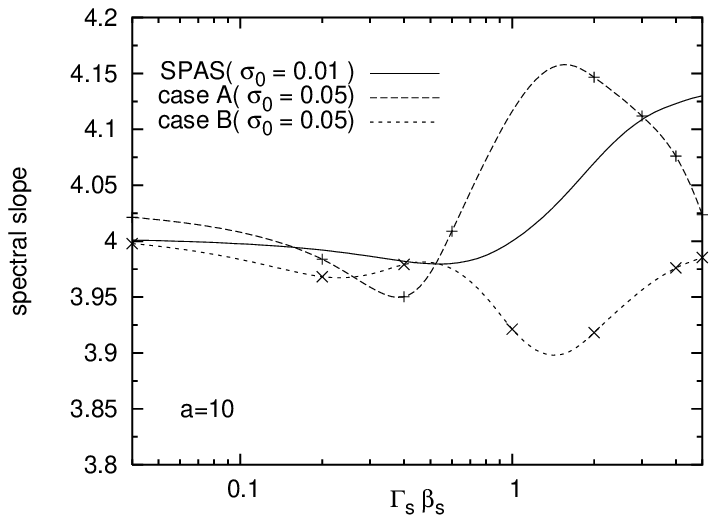} {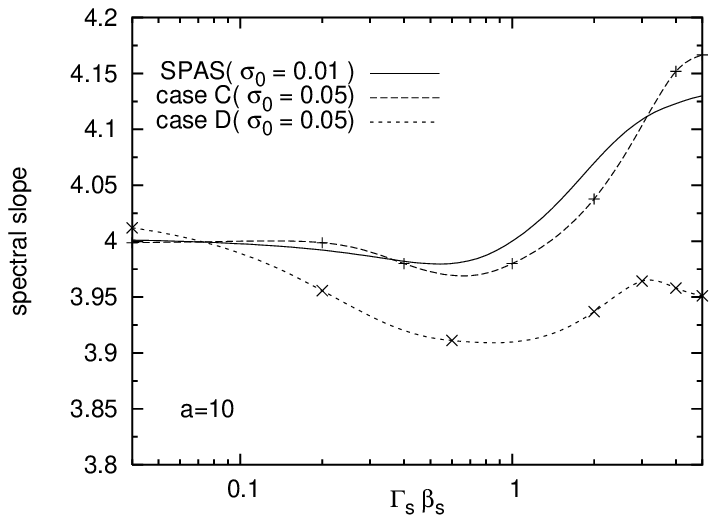}
\caption{Slope vs. shock speed in the four different anisotropic
scattering scenarios: \textit{A} and \textit{B} in the left box, \textit{C} and
D in the right one. All plots are obtained with $a=10$ and $\sigma_0=0.05$.
Slope resulting from isotropic scattering (computed with $\sigma=0.01$) is also shown for
comparison (solid line).
\label{sez3fig_slope_anis_a10}}
\end{center}
\end{figure}

The angular part of the distribution function is shown in Fig.
\ref{sez3fig_ang_dis_anis_AB} for the scenarios \textit{A} (left
panel) and \textit{B} (right panel) and in Fig. \ref{sez3fig_ang_dis_anis_CD}
for the scenarios \textit{C} (left panel) and \textit{D} (right
panel). The slope of the spectrum of accelerated particles is plotted
in Fig. \ref{sez3fig_slope_anis_a10}.

For relativistic shocks, the spread in the slope of the spectrum of
accelerated particles has less spread around $\sim 4$, although in
general it remains true that harder spectra are obtained in the
scenarios \textit{B} and \textit{D}.

\begin{figure}
\begin{center}
\plottwo{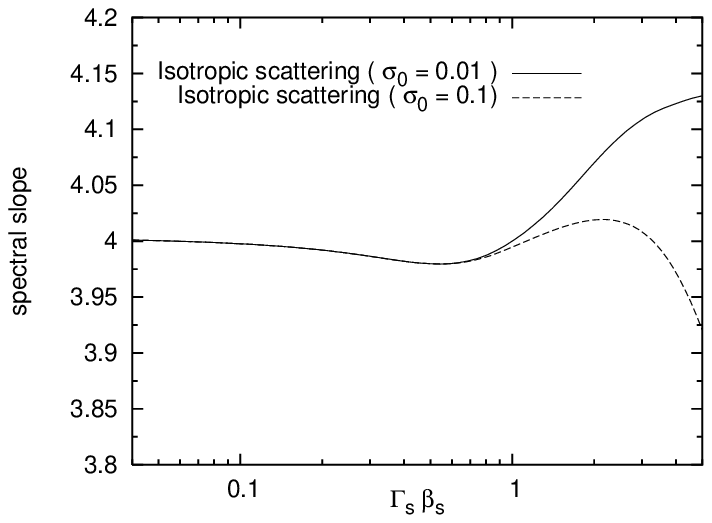} {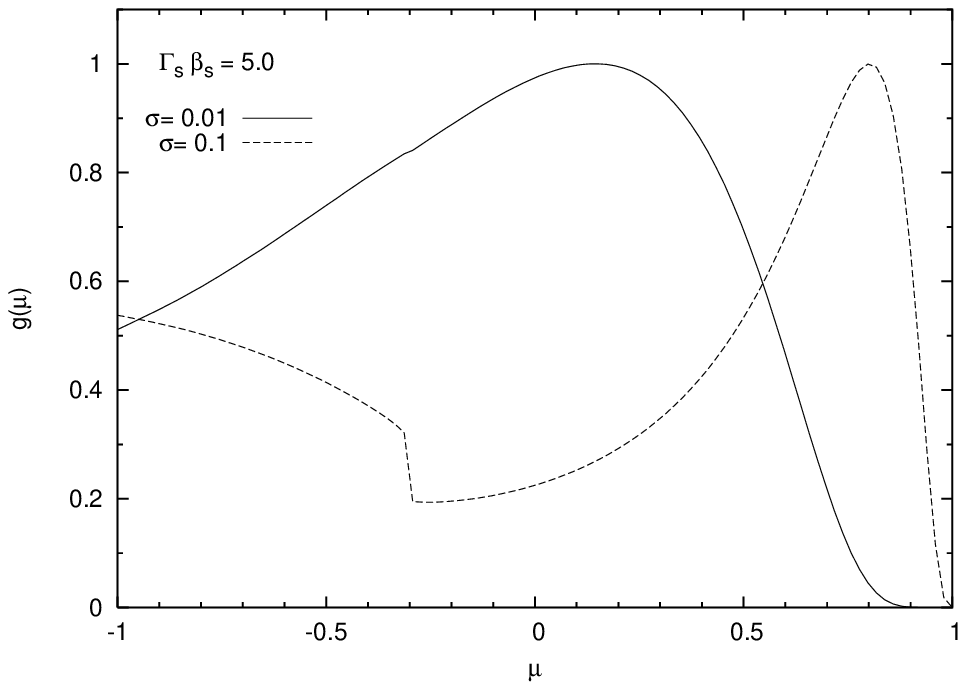} \caption{These plots show the breaking of the
SPAS approximation. \textit{Left box:} slope of the spectrum
of accelerated particles in the case of isotropic scattering with $\sigma=0.01$
(solid line) and $\sigma=0.1$ (dashed line). \textit{Right box:} angular
distribution for $\Gamma_s \beta_s=5$ with $\sigma=0.01$
(solid line) and $\sigma=0.1$ (dashed line).
\label{sez3fig_slope_iso-Break}}
\end{center}
\end{figure}

A note of caution is necessary to interpret the apparent peak in the
slopes at $\Gamma_s \beta_s \sim 1$ for cases \textit{A}, and at
$\Gamma_s \beta_s \sim 3$ for cases \textit{D}. These peaks are completely
unrelated to anisotropic scattering and is instead the result of the breaking of
the regime of small pitch angle scattering (or SPAS), as was already pointed out
in \cite{bla05}. The acceleration process does no longer take place in
the SPAS regime when $\Gamma_s^2 \gtrsim 1/4 \sigma$, which happens at
higher Lorentz factor when $\sigma$ is smaller. This is shown in
Fig. \ref{sez3fig_slope_iso-Break}, where we plot the slope for the
case of isotropic SPAS for $\sigma=0.1$ (dashed line) and
$\sigma=0.01$ (solid line), and the corresponding angular distribution for
$\Gamma_s\beta_s=5$. As already found in \cite{bla05}, the
transition from SPAS to LAS is generally accompanied by a hardening of
the spectra of accelerated particles. The peak seen in
Fig. \ref{sez3fig_slope_anis_a10} is simply the consequence of an
effective value of $\sigma$ in the anisotropic scattering cases
\textit{A} and \textit{D}. This is also clear comparing angular
distributions of Fig. \ref{sez3fig_slope_iso-Break} with the angular
distribution of cases \textit{A} and \textit{D} for $\Gamma_s \beta_s=4$ and 5:
the curves show the same behaviour with a jump at $\mu=-\beta_{\rm down}$ and
a peak that moves towards $\mu=1$ as the shock speed increase.

\section{Conclusions and Discussion}\label{sec:conclusions}

In this paper we carried out exact calculations of the angular
distribution function and spectral slope of the particles accelerated
at plane shock fronts moving with arbitrary velocity, generalizing a
method previously described in detail in \cite{vie03,bla05}. In
particular, we specialized our calculations to two situations: 1)
presence of a large scale coherent magnetic field of arbitrary
orientation with respect to the shock normal, in the upstream
fluid; 2) possibility of anisotropic scattering in the upstream and
downstream plasmas.

Our calculations allowed us to describe the importance of the
inclination of the magnetic field when this has a large coherence
length and there are no scattering agents upstream. For newtonian
shocks, only quasi-perpendicular fields (namely perpendicular to the
shock normal) are of practical importance, in that the return of
particles to the shock from the upstream section is warranted.
Quasi-parallel shocks imply a very low probability of return, so
that the spectrum of accelerated particles is extremely soft. The
process of acceleration eventually shuts off for parallel shocks.
For relativistic shocks, the situation is less pessimistic because the
accelerated particles and the shock front move with comparable
velocities in the upstream frame. In general, the acceleration stops
being efficient when the cosine of the inclination angle $\alpha$ of the
magnetic field with respect to the shock normal is comparable with the
shock speed in units of the speed of light. The slope of the spectrum
of accelerated particles for $\cos\alpha=0$ as a function of the shock
velocity is plotted in Fig. \ref{sez3fig_slope_LAS+B-SPAS+B} for the
two cases in which SPAS or LAS is operating in the downstream plasma.
The slope as a function of $\cos\alpha=0$ for shocks moving at
different speeds is shown in Fig. \ref{sez3fig_slope_cos(alpha)}. In
the same figure we also show the return probability from the upstream
section, in order to emphasize that the presence of a large scale
magnetic field upstream leads to particle leakage to upstream
infinity. This latter phenomenon disappears when scattering is
present, in that scattering always allows for the shock to reach the
accelerated particles. In this case the probability of returning to the
shock at an arbitrary direction is unity. One can ask when and how the
transition from a situation in which there is no scattering to one in
which scattering is at work takes place. When some scattering is
present but the energy density in the scattering agents (e.g. Alfv\`en
waves) is very low compared with the energy density in the background
magnetic field, only very low energy particles are effectively
scattered. When their energy becomes large enough, they only feel the
presence of the coherent field. Increasing the amount of scattering,
this transition energy becomes gradually higher. Particles whose Larmor
radius is larger that the coherence scale of the magnetic field can
eventually escape the accelerator. In general the level of turbulence
(and therefore of scattering) and the number of accelerated
particles are not independent since the turbulence may be
self-generated through streaming-like instabilities \cite{bel78}.

In \S \ref{sec:ani} we extended our analysis to the very interesting
case of anisotropic scattering in both the upstream (unshocked) and
downstream (shocked) medium. The pattern of anisotropy, which clearly
depends on the details of the formation and development of the
scattering centers, has been parametrized in four different
scenarios, and for each one we calculated the angular part of the
distribution function and the spectrum of the accelerated particles.
Deviations from the predictions obtained in the context of isotropic
SPAS and LAS have been quantified: the typical magnitude of these
deflections is a few percent, but there are situations in which the
deviation is more interesting, in particular because it goes in the
direction of making spectra harder.

\acknowledgments
This research was partially funded through grant COFIN 2004-2005. PB
is grateful to the KIPAC at SLAC/Stanford for hospitality during the
final stages of preparation of the manuscript.


\begin{thebibliography}{}

\bibitem[Achterberg {\it et al.} 2001]{ach01}
Achterberg, A., Gallant, Y.A., Kirk, J.G., Guthmann, A.W., 2001, \mnras, 328, 393.

\bibitem[Ballard $\&$ Heavens]{bal91}
Ballard, K.R., Heavens, A.F., 1991, \mnras, 251, 438.

\bibitem[Bednard $\&$ Ostrowski 1998]{bed98}
Bednard, J., Ostrowski, M., 1998, \prl, 80, 3911. \textit{arXiv:astro-ph/9806181 v1}.

\bibitem[Bell 1978]{bel78}
Bell, A.R., 1978, \mnras, 182, 147.

\bibitem[Blasi $\&$ Vietri 2005]{bla05}
Blasi, P., Vietri, M., 2005, \apj, 626, 877.

\bibitem[Freiling, Vietri $\&$ Yurko 2003]{fvy}
Freiling, G., Vietri, M., Yurko, V., 2003, {\it Letters in math.
Phys.}, 64, 65.

\bibitem[Gallant 2002]{gal02}
Gallant, Y.A., 2002, {\it Relativistic flow in Astrophysics}, Lecture Notes in Physics, 589, pp. 24-40.
\textit{arXiv:astro-ph/0302231}.

\bibitem[Gallant $\&$ Achterberg 1999]{gal99}
Gallant, Y.A., Achterberg, A., 1999, \mnras, 305, L6.

\bibitem[Kirk {\it et al.} 2000]{kir00}
Kirk, J.G., Guthmann, A.W., Gallant, Y.A., Achterberg, A., 2000,
\textit{arXiv:astro-ph/000522 v2}.

\bibitem[Kirk $\&$ Duffy 1999]{kir99}
Kirk, J.G., Duffy, P., 1999, J. Phys. G., 25, R163. \textit{arXiv:astro-ph/9905069 v1}.

\bibitem[Kirk $\&$ Schneider 1987]{kir87}
Kirk, J.G., Schneider, P., 1987, \apj, 315, 425

\bibitem[Lemoine $\&$ Pelletier 2003]{lem03}
Lemoine, M., Pelletier, G., 2003, \apj, 589, L73. \textit{arXiv:astro-ph/0304058 v2}.

\bibitem[Lemoine $\&$ Revenu 2005]{lem05}
Lemoine, M., Revenu, B., 2005, \mnras, 000, 1. \textit{arXiv:astr-ph/0510522 v1}.

\bibitem[Niemiec $\&$ Ostrowski 2004]{nie04}
Niemiec, J., Ostrowski, M., 2004, \apj, 610, 851

\bibitem[Synge 1957]{syn57}
Synge, J.L., 1957, {\it The relativistic gas}, North-Holland, Amsterdam.

\bibitem[Peacock 1981]{pea81}
Peacock,J. A., 1981, \mnras, 196, 135.

\bibitem[Vietri 2003]{vie03}
Vietri, M., 2003, \apj, 591, 954.

\end{thebibliography}
\end{document}